\documentclass[trackchanges]{aastex701}

\graphicspath{{./}{figures/}} 
\usepackage{multirow} 
\usepackage{subcaption}
\usepackage{caption}
\usepackage{graphicx}	
\usepackage{amsmath}	
\usepackage{amssymb}	
\usepackage{hyperref}
\usepackage{xcolor}
\usepackage{bm}
\usepackage{booktabs}
\usepackage{hhline}

\usepackage{colortbl}  
\usepackage{xcolor}     
\usepackage{multirow}   
\usepackage{amssymb}   
\usepackage{rotating}  
\usepackage{pifont}    

\newcommand{\cmark}{\textcolor{blue}{\checkmark}}     
\newcommand{\xmark}{\textcolor{lightgray}{\times}}   
     
\usepackage{booktabs} 
\usepackage{multirow}
\usepackage{amsmath, amssymb} 
\usepackage{threeparttable}
\usepackage{booktabs} 
\usepackage{caption}
\usepackage{makecell}

\begin{document}
	

\title{The Deep Learning-Based Dual-Branch Multimodal Fusion Model for Solar Flare Prediction}

\author[sname='Asia China']{Limin Zhao}
\affiliation{School of Space and Earth Sciences, Beihang University, Beijing 100000, China.}
\affiliation{State Key Laboratory of Solar Activity and Space Weather, National Space Science Center, Chinese Academy of Sciences}
\email[show]{limin$\_$zhaom@163.com}  

\author[orcid=0000-0002-1810-6706]{Xingyao Chen} 
\affiliation{State Key Laboratory of Solar Activity and Space Weather, National Space Science Center, Chinese Academy of Sciences}
\email[show]{chenxingyao@nssc.ac.cn}

\author[orcid=0000-0002-1682-1714]{Xiaoshuai Zhu}
\affiliation{State Key Laboratory of Solar Activity and Space Weather, National Space Science Center, Chinese Academy of Sciences}
\email[show]{zhuxiaoshuai@nssc.ac.cn}

\author[orcid=0000-0003-3754-1524]{Dong Zhao}
\affiliation{State Key Laboratory of Solar Activity and Space Weather, National Space Science Center, Chinese Academy of Sciences}
\email[show]{zhaodong@nssc.ac.cn}

\author[orcid=0000-0002-7106-6029]{Yihua Yan} 
\affiliation{School of Space and Earth Sciences, Beihang University, Beijing 100000, China.}
\affiliation{State Key Laboratory of Solar Activity and Space Weather, National Space Science Center, Chinese Academy of Sciences}
\affiliation{School of Astronomy and Space Science, University of Chinese Academy of Sciences, Beijing 100049, China}
\email[show]{yanyihua@nssc.ac.cn}
\correspondingauthor{Yihua Yan}


%
\begin{abstract}
Solar flares are intense eruptive events caused by the rapid release of magnetic energy, often impacting Earth's space environment through electromagnetic radiation and high-energy particles. Accurate flare prediction is critical for space weather forecasting. However, many existing deep learning approaches often rely on single-modal inputs or shallow feature fusion, limiting their ability to capture complementary information. In this study, we propose a dual-branch multimodal fusion deep learning model for predicting 24-hour solar flares. The model integrates magnetograms and magnetic parameters through cross-attention mechanisms, followed by cross-scale interactions at the feature level to enhance multi-scale representation. It is designed to perform both binary prediction of $\geqslant$ C-class flares and multi-class classification of C, M, and X-class flares. To ensure rigorous evaluation, we employ a stratified group five-fold cross-validation scheme to preserve class representativeness and adopt a splitting-before-sampling strategy based on NOAA active region numbers to prevent data leakage. Experimental results show that the model achieves a TSS of 0.661 and an HSS of 0.658 for binary $\geqslant$ C-class prediction, while notably attaining a TSS of 0.780 and an HSS of 0.775 for X-class flares in the multi-class task. Compared with existing approaches, the model demonstrates superior performance in predicting intense X-class flares, effectively suppresses the false alarm rate, and exhibits strong generalization capability.
\end{abstract}

%



\section{Introduction} 
Solar flares are among the most energetic eruptive phenomena in solar activity and represent a rapid and explosive release of magnetic energy stored in the solar corona. This released magnetic energy is converted into electromagnetic radiation across a broad wavelength range and into the acceleration of high-energy particles. 
Intense solar flares are frequently accompanied by the production of solar energetic particles that can propagate through interplanetary space and impact the near-Earth environment \citep{messerotti2009solar}, as well as by coronal mass ejections (CMEs) that expel large magnetized plasma into the heliosphere \citep[e.g.,][]{huang2018deep,zheng2023comparative}.
These eruptive processes can drive rapid and substantial disturbances in the near-Earth space environment, potentially leading to extreme space-weather events \citep{liu2014observations} that disrupt satellite operations, communication and power grids.

Consequently, the development of reliable solar flare forecasting capabilities is essential for mitigating the widespread and potentially severe impacts of solar eruptive events.

The prediction of solar flares is a widely studied topic, with researchers analyzing various prediction methods. \citet{giovanelli1939relations} once proposed a solar flare prediction model based on statistical relationships between solar flares and sunspots. Subsequently, an increasing number of statistical methods are applied to solar flare prediction models \citep[e.g.,][]{ leka2003photospheric,barnes2007probabilistic,song2009statistical,bloomfield2012toward,lee2012solar,falconer2011tool,barnes2016comparison,park2017application}. In addition, there are also many researchers who predict the occurrence of solar flares by studying magnetic field parameters \citep[e.g.,][]{leka2003photospheric,barnes2007probabilistic,yu2009short,huang2010short,korsos2015flare,yang2013magnetic,bobra2015solar,wang2020solar,cicogna2021flare,ribeiro2021machine,yi2021visual,ran2022relationship}. 
For instance, \cite{wang2019parameters} extract SHARP parameters from the polarity inversion line (PIL) region of active regions and employ a support vector machine (SVM) to predict solar flares based on the modified parameters.

Many researchers apply machine learning techniques to solar flare prediction, including artificial neural networks (ANNs) \citep[e.g.,][]{ahmed2013solar,li2013solar,nishizuka2018deep}, support vector machines (SVMs) \citep[e.g.,][]{yang2013magnetic,boucheron2015prediction,bobra2015solar,raboonik2016prediction,nishizuka2017solar,sadykov2017relationships}, random forests (RFs) \citep[e.g.,][]{liu2017predicting,florios2018forecasting}, and multi-layer perceptrons (MLPs) \citep[e.g.,][]{li2013solar,florios2018forecasting}.

Traditional machine learning models require manually designed feature extractors to select and extract relevant features. In contrast, convolutional neural networks (CNNs) are capable of automatically learning and extracting features by training convolutional kernels without human intervention. Consequently, deep learning methods have been increasingly applied to solar flare prediction \citep[e.g.,][]{huang2018deep,park2018application,zheng2019solar,li2020predicting,pandey2024advancing} and long short-term memory (LSTM) networks \citep[e.g.,][]{mccloskey2018flare,chen2019identifying,Liu_2019,jiao2020solar,wang2020predicting,sun2022predicting}.

\citet{huang2018deep} propose a CNN model for binary-class solar flare prediction based on a classical architecture comprising two convolutional layers, each with 64 kernels of size 11×11. \citet{sun2022predicting} train and evaluate two heterogeneous deep learning networks—CNN and LSTM networks—as well as their stacked ensemble. \citet{deshmukh2022decreasing} introduce a hybrid two-stage approach. In the first stage, a CNN model based on the VGG-16 architecture extracts features from a temporal stack of continuous magnetogram samples to generate flare probabilities. These probabilities are then combined with magnetogram-derived feature vectors to train an extremely randomized trees (ERT) model in the second stage, which produces a binary deterministic prediction (flare/no-flare) within the forecast window.

\citet{tang2021solar} propose a solar flare prediction method that fuses three independent deep learning models: deep neural networks (DNNs), bidirectional LSTM networks (Bi-LSTMs), and CNNs. The outputs of these models are integrated using a fully connected neural network to generate the final prediction results. \citet{kaneda2022flare} proposed the Flare Transformer model, which utilizes a magnetogram module and a sunspot feature module to process image and physical features. A transformer attention mechanism is introduced to model temporal relationships between the input features.

In summary, deep learning has made significant breakthroughs and progress in solar flare prediction. Although many CNN-based methods achieve promising results, they still have certain limitations. Some approaches rely solely on magnetograms, others only use physical magnetic parameter features, and some are based on statistical modeling. Although \citet{tang2021solar} attempt to combine the outputs of multiple deep learning models, the overall architecture still adopts a single-branch framework derived from individual networks. In contrast, the Flare Transformer model, proposed by \citet{kaneda2022flare}, employs a sparse attention mechanism that enables only shallow cross-modal interactions, potentially constraining the ability to capture higher-order semantic correlations and limiting effective utilization of magnetic parameters.

Although various deep learning approaches are adopted for solar flare prediction, most existing studies rely on single-branch architectures and shallow fusion strategies, which fail to fully exploit the complementary information between magnetograms and magnetic parameters. Currently, dual-branch multimodal fusion methods based on deep learning remain relatively underexplored for solar flare prediction. In response to these limitations, we propose a dual-branch architecture that encodes magnetograms and magnetic parameters in parallel and employs a cross-modal fusion module at the intermediate stage for deep joint modeling. Unlike traditional methods that merely concatenate features at higher layers or depend on sparse attention for shallow interactions, our approach enables deep fusion during representation learning, effectively capturing complementary features from both modalities. 
In addition, the model incorporates a multi-scale processing mechanism to extract key features across different scales, enhancing its ability to represent complex active regions. By jointly optimizing modality fusion and multi-scale feature representation, our design significantly improves the accuracy and robustness of solar flare prediction.

The structure of the rest sections of the paper is as follows. Section~\ref{sec2:Data Sources} provides a detailed description of the dataset. Section~\ref{sec3:Methods} outlines our proposed methodology and the corresponding models. In Section~\ref{sec4:Experiments}, we evaluate the performance of the proposed solar flare prediction model. Finally, Section~\ref{sec5:Conclusions} concludes the paper with a discussion and summary.

\section{Data Sources} 
\label{sec2:Data Sources}
The SHARP database is derived from full-disk observations by the Helioseismic and Magnetic Imager (HMI) \citep{schou2012design} on the Solar Dynamics Observatory (SDO). It contains the necessary information to automatically identify and track active regions (ARs) in solar magnetograms. Each identified AR is assigned a unique number, and numerous physical parameters relevant to solar flare prediction are provided. Since 2010, these data cutouts have been referred to as Space-weather HMI Active Region Patches (SHARP), which continuously monitor ARs on the solar surface visible to SDO \citep{bobra2014helioseismic}. 

In this study, we conduct an experimental analysis of solar flare prediction using SHARP data provided by the SDO/HMI instrument team. All magnetograms of recorded ARs from May 2010 to May 2023, with a cadence of 96 minutes, and in the Cylindrical Equal-Area (CEA) projection are selected. 

In addition, based on the $ \text{X} $-ray irradiance data recorded by the Geostationary Operational Environmental Satellites (GOES), operated by the National Oceanic and Atmospheric Administration (NOAA), information on the location, intensity, and timing (start, peak, and end) of solar flares is obtained. Solar flares are classified into five categories—A, B, C, M, and X—according to their peak soft $ \text{X} $-ray flux. The A and B classes are generally considered minor solar flares, whereas C, M, and X-class flares are classified as major solar flare events, with M and X-class flares being the most intense and thus of particular importance.

\subsection{Feature Set of Physical Parameters}
\begin{table}
	\centering
	\caption{ List of SHARP parameters used in this study and a brief description of the parameters. }
	\label{tab01:feature_parameter}
	\renewcommand{\arraystretch}{1.1}
	\begin{tabular}{lc} 
		\toprule  
		Keyword  & \qquad Description \\
		\midrule 
		TOTUSJH  & \qquad Total unsigned current helicity \\
		TOTPOT   & \qquad Total photospheric magnetic free energy density \\
		TOTUSJZ  & \qquad Total unsigned vertical current \\
		ABSNJZH  & \qquad Absolute value of the net current helicity \\
		SAVNCPP  & \qquad Sum of the modulus of the net current per polarity \\
		USFLUX   & \qquad Total unsigned flux \\
		AREA\_ACR  & \qquad Area of strong field pixels in the active region \\
		MEANPOT  & \qquad Mean photospheric magnetic free energy \\
		R\_VALUE & \qquad Sum of flux near lolarity inversion line(PIL)  \\
		SHRGT45  & \qquad Fraction of area with shear $ > 45^\circ $ \\
		\bottomrule
	\end{tabular}
\end{table}
Each record in the SHARP dataset includes a set of metadata containing values of various physical magnetic parameters for the corresponding AR. These parameters are primarily derived from raw magnetic flux observations and have been continuously updated and refined. They are widely regarded as informative and physically meaningful indicators for solar flare prediction.

A number of studies have investigated photospheric magnetic field parameters to identify those most relevant to solar flare activity.
\citet{leka2003photospheric} find that parameters associated with free magnetic energy, total unsigned flux, vertical current, and shear angle are strongly correlated with flare occurrence. Similarly, \citet{ahmed2013solar} show that using only six magnetic parameters yields predictive performance comparable to that obtained with all 21 features.

\citet{bobra2015solar} systematically evaluated 25 magnetic parameters provided in the SHARP data products and identified several key features, such as TOTUSJH, TOTPOT, and TOTUSJZ, that are particularly effective in distinguishing flaring from non-flaring active regions, consistent with the findings of \citet{leka2003photospheric}. Building on this work, \citet{liu2017predicting} select 13 SHARP parameters for solar flare prediction and further demonstrate that TOTUSJZ, TOTUSJH, and R\_VALUE are especially important for distinguishing flares of different classes. Subsequently, \citet{ran2022relationship} analyze that several SHARP parameters, including SAVNCPP, MEANPOT, USFLUX, SHRGT45, and ABSNJZH exhibit strong correlations with flare occurrence.

In this study, we adopt the SHARP magnetic field parameters listed in Table~\ref{tab01:feature_parameter} as the input feature set for solar flare prediction. These parameters are selected because they are important for flare prediction and have been frequently used in recent studies\citep[e.g.,][]{liu2017predicting, tang2021solar,zheng2023multiclass,zheng2023comparative,li2024prediction}.

\subsection{Dataset Preprocessing}
Despite the high quality of the SHARP data products, the raw dataset inevitably contains observational noise, missing values, and low-quality samples. To ensure data reliability, we adopt a strictly cascaded filtering strategy to construct a high-quality experimental dataset.

In the data cleaning phase, we first discard records lacking NOAA active region number and retain only high-quality samples based on the official SDO/HMI Quality Flags. Considering that regions near the solar limb are susceptible to projection effects, we select only active regions located within $\pm 70^{\circ}$ of the central meridian. 
Furthermore, to minimize irrelevant noise and exclude ARs with insufficient scale or information, we implement an adaptive screening strategy based on the solar cycle. Specifically, we independently enforce constraints on both physical scale and image resolution within each cycle: regions falling in the bottom 5\% of the area distribution are first removed; subsequently, only samples with a pixel count exceeding 1000 are retained to ensure adequate spatial resolution. Simultaneously, all records containing missing values or physical anomalies are eliminated.

In the preprocessing phase, statistical analysis reveals that key SHARP parameters (e.g., total magnetic flux, current helicity) exhibit significant scale disparities. Using raw values directly can lead to training instability. Therefore, we apply a logarithmic transformation with a unit offset to highly skewed parameters, followed by a unified Z-score standardization for all physical parameters. This process effectively eliminates scale differences and ensures numerical stability during multimodal feature fusion.

\subsection{Labeling and Sampling Process}
\label{sec2.3:sample_label}
After completing feature selection and data preprocessing, a standardized dataset of magnetogram samples is constructed. In this section, we detail the construction of the datasets used for model training and evaluation. 

\subsubsection{Sample Labeling}
\begin{table}
	\centering
	\caption{ List of dataset used in this study. }
	\label{tab02:dataset_details}
	\renewcommand{\arraystretch}{1.0}
	\begin{tabular}{cccc} 
		\toprule  
		Category & \; Dataset&\;Number& \;Forecast Window\\
		\midrule	
		\multirow{2}{*}{Two}&\; no-flare & \;99,059 &\quad\multirow{2}{*}{24h} \\
		& Flare (C,M,X) & \; 35,191  \\
		\hline
		\multirow{3}{*}{Three}
		&C flare & \;29,327 &\quad \multirow{3}{*}{24h} \\
		&M flare & \;5,337  \\
		&X flare & \;527  \\
		\bottomrule
	\end{tabular}
\end{table}
Due to the requirement for labeled training samples in deep learning, we utilize the GOES $ \text{X} $-ray Flare Catalog provided by the National Centers for Environmental Information (NCEI)\footnote{\url{https://hesperia.gsfc.nasa.gov/goes/goes_event_listings/}}. Since the NCEI catalog omits AR information for certain events, we implement a stepwise supplementation process. 

Specifically, we initially cross-match records lacking AR identification with the SolarSoft event archive\footnote{\url{https://www.lmsal.com/solarsoft/latest_events_archive.html}} to supplement the missing AR numbers. Subsequently, for events still missing AR associations, we manually inspect the corresponding records from SDO\footnote{\url{https://www.solarmonitor.org/}}. 
After this procedure, 3,440 flare records are successfully supplemented. Consequently, the proportion of records lacking AR information is reduced to only 3.1\% (649 out of 21,067), which are excluded from the subsequent analysis. The supplemented flare catalog is publicly available on Flare-Catalog \footnote{\url{https://github.com/zlm163com/Flare-Catalog}}.

After a series of preprocessing steps—including the removal of samples with missing ARs, elimination of records with incomplete metadata, and normalization of magnetic parameters—a total of 134,250 magnetogram samples are constructed. This dataset comprises not only visual features from magnetograms but also features derived from their corresponding magnetic parameters, which are of critical importance for analyzing and predicting solar flare activity.

Solar flare prediction is treated as a classification task here. Building on the revised catalog, the study aims to predict whether a C, M, or X-class solar flare occurs in a given active region within the next 24 hours, focusing on events that may significantly impact Earth and human activities.

For the binary task of predicting whether a flare occurs, the sample is labeled accordingly. Specifically, based on the records from NCEI, if a C-class or stronger flare occurs within 24 hours following a given timestamp, the corresponding magnetogram sample is labeled as positive; otherwise, it is labeled as negative. The resulting dataset consists of 35,191 positive samples and 99,059 negative samples.

Building upon the binary classification task, the fine-grained prediction task further labels each positive sample according to the most intense flare that occurred within the 24-hour window following the timestamp. Specifically, samples are assigned to three classes: C-class, M-class, and X-class flares. The resulting dataset includes 29,327 C-class samples, 5,337 M-class samples, and 527 X-class samples. The detailed distribution is shown in Table~\ref{tab02:dataset_details}.

\subsubsection{Data Sampling and Splitting Strategy} 
Solar flare prediction datasets inherently exhibit severe class imbalance and strong spatiotemporal correlations. To ensure a rigorous evaluation, we design a comprehensive data processing pipeline that strictly enforces a splitting-before-sampling sequence implemented via AR-based grouping.

A critical step in prediction is dividing the dataset into training and testing sets. Traditional random splitting leads to information leakage due to highly correlated samples from the same AR, while year-based partitioning suffers from solar cycle variations that can result in an absence of major flares in certain testing periods. To overcome these limitations, we adopt a strict AR-based splitting strategy. All samples from a unique AR number are assigned exclusively to either the training or testing set, ensuring physical isolation between the subsets.

Beyond physical isolation, the sequence of data processing is critical. A common methodological pitfall is applying sampling strategies to the entire dataset prior to splitting. Even when AR-based splitting is subsequently applied, this procedure introduces statistical leakage, as the construction of the training set becomes implicitly coupled with the distribution of the testing set. This violates the fundamental principle that the training pipeline should remain independent of unseen test data.

To address this issue, we strictly adopt a splitting-before-sampling strategy. Under this approach, all samples from a given AR are assigned exclusively to either the training or testing set, with sampling operations applied independently to each subset only after the split is finalized. This approach offers two advantages: (1) it ensures statistical independence by keeping the sampling process entirely separate from the test data, thereby preventing any implicit information leakage; and (2) it prevents the model from memorizing region-specific evolutionary signatures, thereby providing a more reliable assessment of its true generalization capability.

\begin{table} 
	\centering
	\caption{ List of binary prediction datasets.}
	\label{tab03:two_data}
	\renewcommand{\arraystretch}{1.0}
	\setlength{\tabcolsep}{1.8mm}{
		\begin{tabular}{ccccc}
			\toprule
			\multirow{2}{*}{Datasets} & \multirow{2}{*}{-} & \multicolumn{1}{c}{\multirow{2}{*}{\shortstack{Forecast \\ Window}}} & \multirow{2}{*}{No-Flare} & \multirow{2}{*}{$ \geqslant $C-class} \\
			& & & & \\
			\midrule	
			\multirow{2}{*}{D1}&Training&\multirow{2}{*}{24h}&79146&27643\\
			&Testing&&19913&7548\\	
			\multirow{2}{*}{D2}&Training&\multirow{2}{*}{24h}&47488&46993\\
			&Testing&&11948&12832\\
			\bottomrule
		\end{tabular}
	}
\end{table}

To mitigate the severe class imbalance within the dataset, we employ a hybrid sampling strategy. For the majority class, we apply random undersampling to reduce data redundancy and computational cost. Conversely, for the minority class, we implement geometric data augmentation. These minority-class magnetogram samples undergo discrete rotations and random horizontal or vertical flipping. These transformations exploit the inherent rotational invariance of active regions, effectively expanding the feature space of rare events while encouraging the model to learn robust, orientation-independent features. This data strategy is consistent with methodologies widely adopted in previous solar flare prediction studies \citep[e.g.,][]{zheng2019solar, zheng2021hybrid}.

Based on these principles, specific datasets are constructed for binary and multi-class prediction tasks:

For the binary prediction task targeting flares of C-class and above, we implement a stratified group hold-out strategy with an 8:2 training-testing ratio based on active regions. Based on this splitting, we establish two evaluation environments. Dataset D1 preserves the original imbalanced observational distribution, while Dataset D2 is constructed by applying our hybrid sampling techniques independently to the subsets after the initial split. Crucially, all sampling for D2 is performed independently within each subset after the split to ensure statistical isolation. The detailed class distributions for D1 and D2 are listed in Table \ref{tab03:two_data}.

\begin{table*} 
	\centering
	\caption{ List of stratified group five-fold cross-validation datasets.}
	\label{tab04:k-fold}
	\renewcommand{\arraystretch}{1.05}
	\setlength{\tabcolsep}{4.3mm}{
		\begin{tabular}{lccccc} 
			\toprule
			\multirow{2}{*}{Fold} & \multirow{2}{*}{-} & \multicolumn{1}{c}{\multirow{2}{*}{\shortstack{Forecast \\ Window}}} & \multirow{2}{*}{C}& \multirow{2}{*}{M} &\multirow{2}{*}{X} \\
			&&&&& \\
			\midrule	
			\multirow{2}{*}{F1}&Training&\multirow{10}{*}{24h}&4663&5081&4300 \\
			 &Testing&&1202&1324&1261 \\
		    \multirow{2}{*}{F2}&Training&&4762&5143&4410 \\
	         &Testing&&1104&1261&1118\\
			\multirow{2}{*}{F3}&Training&&4607&5225&4010 \\
			&Testing&&1258&1180&1260 \\
			\multirow{2}{*}{F4}&Training&&4858&4925&4230 \\
			&Testing&&1008&1480&1040 \\
			\multirow{2}{*}{F5}&Training&&4571&5244&4130 \\
			&Testing&&1294&1160&1140 \\
			\bottomrule
		\end{tabular} }
\end{table*}

Building upon the binary framework, we further extend the analysis to a fine-grained multi-class prediction task to evaluate the model’s ability to discriminate among specific flare levels. However, given the extreme scarcity of M and X-class flares, reliance on a single hold-out split may yield unreliable results and introduce substantial evaluation bias. In extreme cases, this can even result in the complete absence of rare flare classes in the testing set. 
To ensure evaluation robustness, we adopt a stratified group five-fold cross-validation scheme, splitting the entire dataset into five mutually exclusive folds based on active regions, denoted as F1–F5. 
In this strategy, each fold sequentially serves as the testing set while the remaining four form the training set. This stratified approach explicitly preserves the relative proportions of C, M, and X-class flares across all folds, ensuring that rare events are adequately represented. Consistent with our protocol, all class-balancing operations are performed independently within each fold after the splitting is finalized to prevent information leakage. The resulting distributions for each fold are detailed in Table \ref{tab04:k-fold}.

\section{Methods}
\label{sec3:Methods} 
This study proposes a transformer-based multimodal information fusion architecture for solar flare prediction. The model employs a parallel dual-branch transformer structure to enable cross-scale feature extraction and fuses multimodal information from solar magnetograms (image modality) and magnetic parameters (text modality) to predict solar flare activity within the next 24 hours.

\begin{figure*}
	\includegraphics[width=\textwidth]{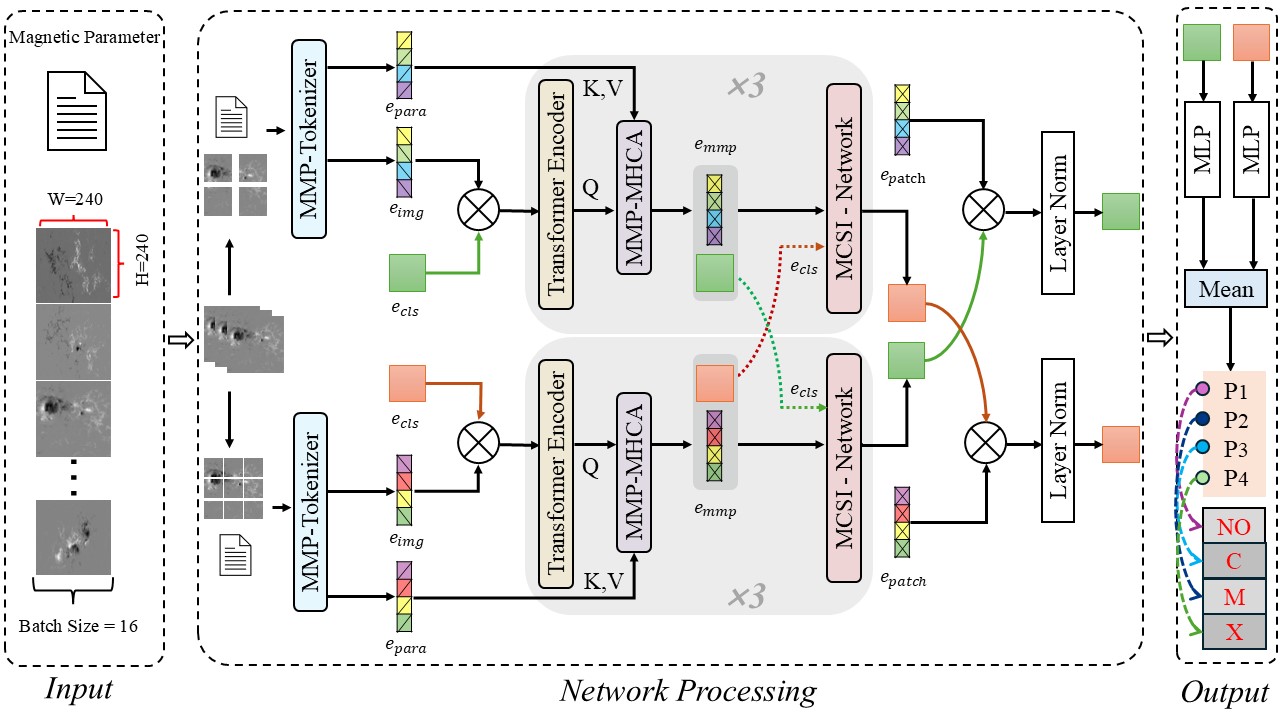}
	\caption{ The proposed network architecture consists of three main components: the Magnetogram and Magnetic Parameter Tokenizer (MMP-tokenizer), the Multi-Head Cross Attention module (MMP-MHCA), and the Multimodal Cross-Scale Interaction Network (MCSI-network). It incorporates $ M=3 $ multimodal transformer encoders, each containing two branches processing magnetogram patches at different scales. Multimodal fusion with the corresponding magnetic parameter tokens is performed via an attention mechanism in each branch. Cross-scale information fusion is achieved by using the CLS tokens from each branch. The extracted feature vectors from magnetogram patches and magnetic parameters are denoted as $\bm{e}_{\mathrm{img}}$ and $\bm{e}_{\mathrm{para}}$, respectively. Their fused multimodal representation is denoted as $\bm{e}_{\mathrm{mmp}}$. Here, $\bm{e}$ represents a feature vector, and the subscript indicates its corresponding modality. }
	\label{fig01:1_net}
\end{figure*}

\subsection{Backbone}
As illustrated in Figure~\ref{fig01:1_net}, the proposed architecture is a dual-branch transformer designed to predict solar flares from multi-modal inputs. The backbone consists of three core integrated modules:

Magnetogram and Magnetic Parameter Tokenizer (MMP-Tokenizer): Located at the input stage, this module transforms the heterogeneous inputs into unified token embeddings. It specifically integrates CNN-extracted visual features from magnetograms with the projected physical parameters, mapping these different modalities into a shared feature space for subsequent processing.

Magnetogram and Magnetic Parameter Multi-Head Cross-Attention (MMP-MHCA): Embedded within the transformer encoder of each branch, this module performs deep multimodal fusion. It enables dynamic interaction between visual features and magnetic parameters to compensate for the limitations of single-modal data, ensuring the capture of complementary information from both modalities.

Multimodal Cross-Scale Interaction Network (MCSI-Network): To obtain a comprehensive multi-scale representation, this module employs a cross-attention mechanism to fuse features from the parallel branches. It explicitly integrates fine-grained local details with coarse-grained global features, producing a robust embedding that combines the strengths of both spatial views.

\subsection{MMP-Tokenizer}
To enable transformer-based modeling of heterogeneous data, we propose a unified tokenization module termed MMP-Tokenizer. In our model, this module serves as a modality interface. Its primary function is to project both high-dimensional magnetograms and their associated scalar magnetic parameters into token representations of a consistent dimension ($D$). This alignment allows the subsequent Transformer to perform cross-modal attention and fusion within a shared embedding space.

\begin{figure*} 
	\includegraphics[width=\textwidth]{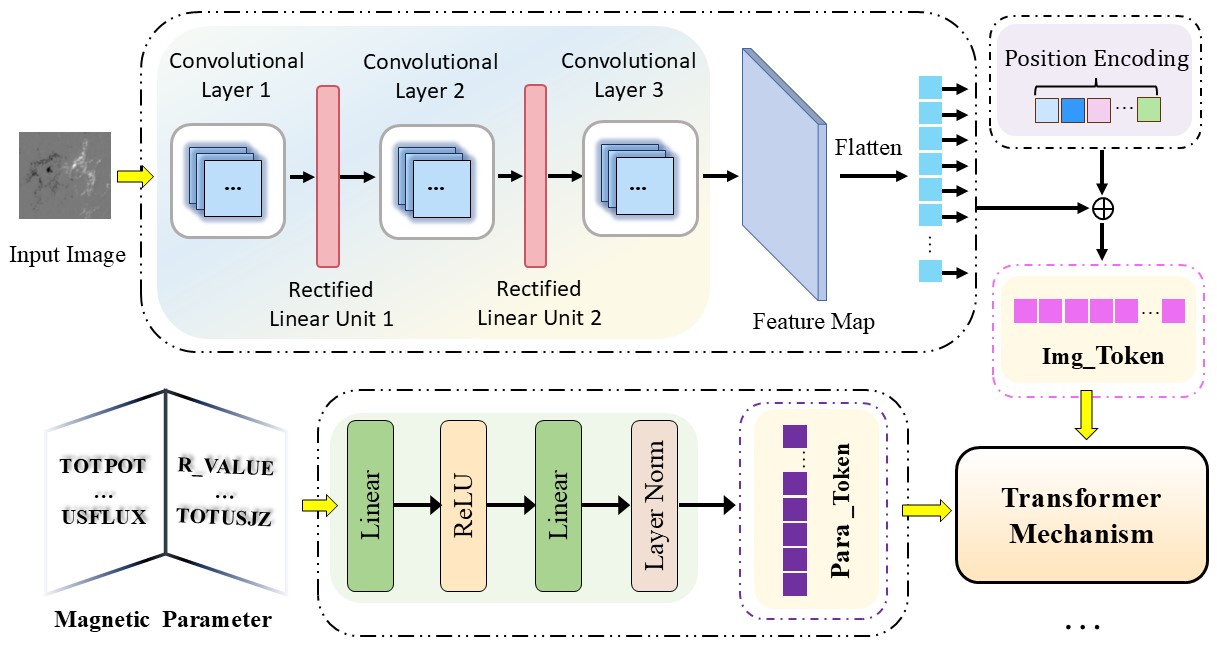}
	\caption{ The CNN module for extracting features from magnetograms and the structure for converting magnetic parameters into token representations. }
	\label{fig02:2_CNN}
\end{figure*}

As illustrated in Figure~\ref{fig02:2_CNN}, the module consists of two parallel branches. The top branch processes magnetograms into visual tokens using a convolutional network, while the bottom branch transforms magnetic parameters into token embeddings using a feedforward network.

After obtaining token representations from both branches, these tokens are then utilized at different stages of the model: visual tokens are fed into a transformer encoder for high-level spatial representation learning, while parameter tokens are subsequently fused with encoded visual features through a multimodal fusion module. 

For the visual branch, to enhance the representation of magnetogram features, this study adopts a convolution-based patch embedding strategy inspired by the Vision Transformer (ViT) framework \citep{dosovitskiy2021image}. Specifically, we replace the standard linear patch projection layer commonly used in ViT architectures with a three-layer convolutional network.
Given an input image $I$, the network first employs a $7 \times 7$ convolutional layer to capture coarse-grained structural features. This is followed by a $3 \times 3$ convolutional layer to extract fine-grained local details. A final $3 \times 3$ convolutional layer projects the output channels to match the target embedding dimension ($D$). ReLU activation functions are applied after the first two convolutional layers to introduce non-linearity. Finally, the resulting feature map is reshaped into a sequence of visual tokens only after these spatial features are captured, providing a structured input that can be directly fed into the transformer encoder.

The feature extraction process can be formally described as:
\begin{equation}
\begin{aligned}
F_1 &= \sigma(W_1 * I + b_1), \\
F_2 &= \sigma(W_2 * F_1 + b_2), \\
F_3 &= W_3 * F_2 + b_3, \\
T &= \text{Reshape}(F_3) \in \mathbb{R}^{ B \times N \times D}.
\end{aligned}
\end{equation}
Here, \( I \in \mathbb{R}^{B \times C \times H \times W} \) denotes the input magnetogram, where $B$ is the batch size, $C$ is the number of input channels, and $H$, $W$ represent the height and width of the image, respectively. The symbol $ \sigma  $ denotes the ReLU activation function, and $W_{i} $ and $ b_{i} $ represent convolutional weights and biases. The reshape operation flattens the spatial feature map $F_3$ into a sequence of $N$ tokens, each of dimension $D$, which are directly fed into the transformer encoder.

This convolution-based tokenization strategy offers three main advantages. First, the local connectivity of convolutions helps preserve fine-grained features in magnetograms—such as local gradients and texture patterns—thus improving the quality of spatial representations. Second, the hierarchical structure of convolutional layers progressively expands the receptive field, allowing the network to capture multi-scale features that enrich subsequent multimodal fusion. Finally, the inductive bias introduced by CNNs enhances the model’s generalization and stabilizes transformer optimization, particularly when training data is limited. Indeed, our empirical analysis confirms that replacing this module with a standard linear projection significantly reduces performance, showing the critical role of convolutional feature extraction in capturing local magnetic patterns. Simultaneously, magnetic parameters are processed in parallel through the parameter tokenization branch, described below.

In parallel with the CNN-based magnetogram tokenizer, the magnetic parameters are transformed into token representations. Let $P \in \mathbb{R}^{B \times N_{p}}$ denote the input parameter value matrix, where $ B $ is the batch size and $ N_{p} $ is the number of magnetic parameters. To enable embedding of parameter values, the input is reshaped to $ \mathbb{R}^{B \times N_{p} \times 1 } $, such that each parameter value can be individually encoded. The reshaped tensor is then passed through a lightweight feedforward network. Non-linearity is introduced through ReLU activation, and a LayerNorm layer is applied at the end to stabilize the feature distribution. This results in a tensor of parameter tokens $ T_{p} \in \mathbb{R}^{B \times N_{p} \times D} $, where $ D $ denotes the target embedding dimension consistent with the visual token space.

This tokenization strategy ensures that the parameter tokens are architecturally aligned with visual tokens, enabling effective multimodal interaction in the subsequent fusion stage.

\subsection{Vision Transformer}
The CNN transforms the magnetogram into a sequence of $ D $-dimensional visual tokens, each corresponding to a local region of the image. These tokens are then combined with positional encodings to retain spatial information and are subsequently fed into the transformer encoder for high-level representation learning.

\begin{figure}
	\centering
	\includegraphics[width=0.8\columnwidth]{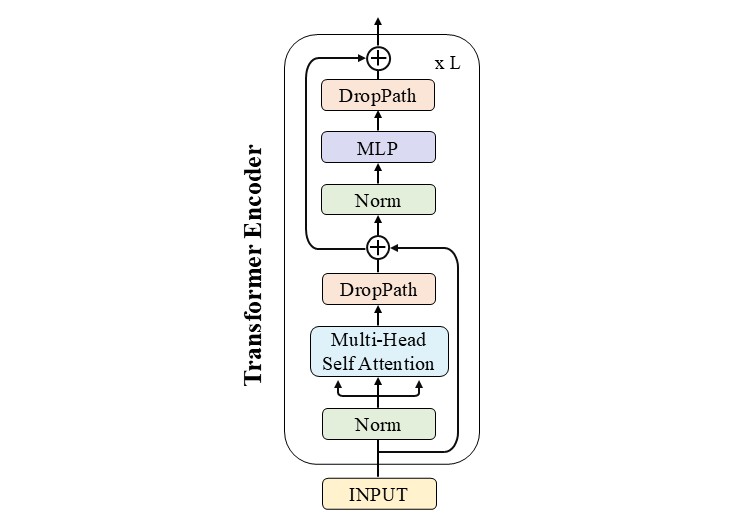} 
	\caption{Architecture of the multi-branch transformer encoder module. }
	\label{fig03:3_block}
\end{figure}
As shown in Figure~\ref{fig03:3_block}, the transformer encoder processes the sequence of tokens through multiple blocks. Each block includes layer normalization, multi-head self-attention, and a feedforward network, along with residual connections that facilitate feature propagation. This architecture enables the model to capture high-level spatial representations across the entire magnetogram. Finally, the high-level visual features extracted from the transformer are fused with magnetic parameters to enhance the model’s predictive capability. 

\subsection{Cross-Attention Fusion}
To effectively integrate information from both magnetograms and magnetic parameters, we propose a multimodal attention fusion module, termed MMP-MHCA.
Drawing inspiration from the foundational attention architecture \citep{vaswani2017attention} and its successful extensions in multimodal learning \citep[e.g.,][]{lu2019vilbert, li2021align}, our MMP-MHCA module employs a multi-head cross-attention mechanism.

This design leverages the mechanism's inherent ability to fuse heterogeneous features by allowing the query to originate from the visual modality (magnetogram) while the keys and values are derived from the physical parameter modality. This formulation enables fine-grained semantic interaction where magnetic parameters dictate attention-based weighting over the spatial tokens of the magnetograms, thereby allowing the model to selectively emphasize semantically relevant regions for effective feature enrichment.

During the multimodal fusion process, the magnetogram features and magnetic parameter features are denoted as $ X_v \in \mathbb{R}^{B \times N \times D} $ and $ X_{t} \in \mathbb{R}^{B \times N_p \times D} $ respectively, where $ B $ is the batch size, $ N $ is the number of visual tokens, $N_p$ is the number of parameter tokens, and $ D $ is the feature dimension. Here, $ X_v $ serves as the query (Q), and $ X_t $ serves as both the key (K) and the value (V). Before computing attention weights via dot-product operations, the input features are linearly projected to obtain the Q , K, and V matrices.
\begin{equation} 
Q = X_v W_q, \quad K = X_t W_k, \quad V = X_t W_v 
\end{equation}
where $ W_{q} $, $ W_{k} $ , $ W_{v} $ are the weight matrices for \textbf{\( Q \)}, \textbf{\( K \)}, and \textbf{\( V \)}, respectively.

In the multi-head attention mechanism, each head independently learns linear projections to produce Q, K, and V matrices. Specifically, for the i-th attention head:
\begin{equation} 
Q^{(i)} =X_{v}W_{q}^{(i)},\quad K^{(i)} = X_{t}W_{k}^{(i)}, \quad V^{(i)} = X_{t}W_{v}^{(i)}
\end{equation}
where  $W_q^{(i)}$, $W_k^{(i)}$, and $ W_v^{(i)} \in \mathbb{R}^{D \times d_{head}}$ denote the weight matrices for the $i$-th head, and $d_{head} = D / N_{head}$ is the dimension per head.

Mathematically, the cross-attention weight for each head can be formulated as:  
\begin{equation}
A^{(i)} = \text{softmax}\left(\frac{Q^{(i)} {K^{(i)}}^\top}{\sqrt{d_{head}}}\right),
\end{equation}
where \( A^{(i)} \) represents the normalized attention weight matrix for the $ i $-th head, capturing the relevance between each spatial Q token and all key positions from the magnetic parameters.

The output of the $ i $-th attention head, denoted as $ Z^{(i)} $, is computed via scaled dot-product attention as
\begin{equation}
Z^{(i)} = A^{(i)} V^{(i)} \in \mathbb{R}^{B \times N \times d_{head}},\\
\end{equation}
Each of the $ N $ spatial tokens in $ Z^{(i)} $ aggregates contextually relevant information based on these attention weights.

Finally, the outputs of all attention heads are concatenated and passed through a linear transformation to generate the final fused representation:
\begin{equation}
\bm{e}_{mmp} = \text{concat}(Z^{(1)}, \dots, Z^{(N_{head})}) W_{o},
\end{equation}
where \( W_o \) denotes the output projection matrix.

\subsection{MultiModal Cross-Scale Encoder}
The Vision Transformer (ViT) \citep{dosovitskiy2021image} framework treats images as sequences of patches. Adopting the BERT-style~\citep{devlin2019bert} sequence modeling paradigm, ViT prepends a learnable class token (CLS) to the input sequence within its bidirectional Transformer architecture. Through the standard self-attention mechanism, this CLS token gathers information from all image patches to construct a compact and discriminative global feature embedding, effectively serving as a global semantic representation.

In our proposed framework, given that multiscale images provide complementary spatial granularities, we employ two parallel branches to process features at different scales. Each branch adopts a similar architecture; the large-scale (L) branch processes larger image patches, while the small-scale (S) branch handles smaller image patches. Within each branch, as the magnetogram passes through the encoder layers, the CLS token first aggregates the specific visual features of that scale. In the subsequent MMP-MHCA module, the CLS token serves as the query, while the magnetic parameters are used as keys and values to enable cross-modal interaction. 
Through this process, each branch yields the global Multimodal Embedded Magnetogram and Magnetic Parameters, denoted as $\bm{e}^i_{\mathrm{mmp}}$ (where $i \in \{l, s\}$). This representation comprises the updated multimodal CLS token $\bm{e}^i_{\mathrm{cls}}$ and the spatial patch tokens $\bm{e}^i_{\mathrm{patch}}$. 

Although the multimodal representation $\bm{e}_{\mathrm{mmp}}$ provides rich information at each specific scale, relying solely on single-scale features may limit the model’s ability to capture global context across different spatial resolutions. 
To address this limitation, drawing inspiration from the success of multi-scale architectures such as HRNet \citep{sun2019deep}, as well as recent dual-branch transformers \citep[e.g.,][]{lee2022mpvit,havtorn2023msvit,yao2023dual,chen2025dual,chen2021crossvit}, we introduce the Multimodal Cross-Scale Interaction Network (MCSI-Network).
It employs an attention-based token interaction strategy to fuse multi-scale CLS and $\bm{e}_{\mathrm{patch}}$ representations.

In the subsequent cross-scale interaction stage, $\bm{e}_{\mathrm{cls}}$ serves as the query to interact with $\bm{e}_{\mathrm{patch}}$ from the other branch. Specifically, we first slice the feature vectors from the multimodal sequences $\bm{e}_{\mathrm{mmp}}$ based on token positions. The token at the 0-th position is extracted as the multimodal CLS token $\bm{e}_{\mathrm{cls}}$, while the subsequent tokens constitute the spatial patch sequence $\bm{e}_{\mathrm{patch}}$.

\begin{figure} 
	\centering
	\includegraphics[width=0.6\columnwidth]{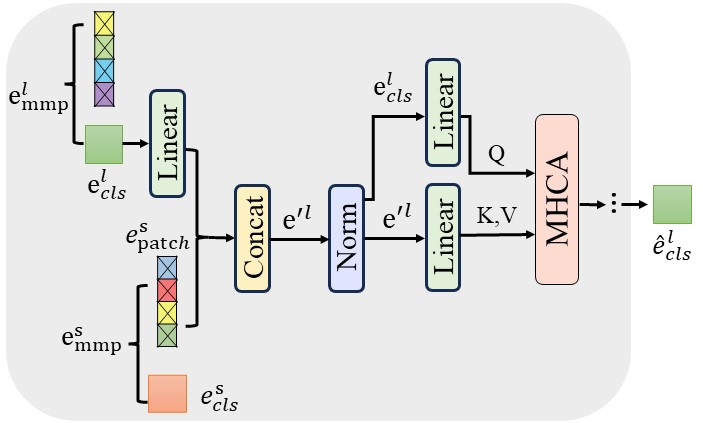}  
	\caption{ Cross-attention information fusion between different branches. }
	\label{fig04:4_scale}
\end{figure}
As illustrated in Figure~\ref{fig04:4_scale}, we take the update of the L branch as an example. The CLS token $\bm{e}_{\mathrm{cls}}^{l}$ from the L branch is first projected and then concatenated with the patch tokens $\bm{e}_{\mathrm{patch}}^{s}$ from the S branch. This forms a hybrid sequence $\bm{e}'^{l}$, which is input into the multi-head cross-attention (MHCA) module. Then, $\bm{e}'^{l}$ serves as the keys and values to interact with the query $\bm{e}^l_{\mathrm{cls}}$. Finally, a residual connection is applied to stabilize the training and effectively integrate the cross-scale features. This process is formalized as:
\begin{equation}
	\begin{gathered}
		\bm{e}^l_{\mathrm{cls}} = \bm{e}^l_{\mathrm{mmp}}[0], \quad \quad
		\bm{e}^s_{\mathrm{patch}} = \bm{e}^s_{\mathrm{mmp}}[1{:}N] \\[2ex] 
		\bm{e}'^{l} = \text{Concat} \left( f^l\left( \bm{e}_{\mathrm{cls}}^l \right),\  \bm{e}_{\mathrm{patch}}^s \right) \\[2ex]
		\bm{\hat{e}}^l_{\mathrm{cls}} = \bm{e}^l_{\mathrm{cls}} + \text{MHCA}\left(Q = \bm{e}^l_{\mathrm{cls}} W_q, \ K = \bm{e}'^{l} W_k, \ V = \bm{e}'^{l} W_v\right).
	\end{gathered}
\end{equation}
where $N$ denotes the patch sequence length; $W_q, W_k, W_v$ are learnable projection parameters; and $f^l(\cdot)$ denotes the projection function used for dimensional alignment. The small-scale branch acquires its corresponding updated CLS token $\bm{\hat{e}}_{\mathrm{cls}}^{s}$ in a symmetric manner.

Subsequently, the updated multimodal cross-scale CLS token is reinjected into its original position to reconstruct the sequence, preserving the structural integrity of the branch.
\begin{equation}
\bm{e}^{i}_{\mathrm{out}}=\text{Concat}(\bm{\hat{e}}^{i}_{\mathrm{cls}},\  \bm{e}^{i}_{\mathrm{patch}}).
\end{equation}
After processing by the MCSI-Network, the sequences effectively fuse multimodal and multiscale information and are utilized for the final solar flare prediction.

\section{Experiments}
\label{sec4:Experiments}
In this section, we present the experimental evaluation of the proposed model using magnetograms and corresponding magnetic parameters as inputs. Based on the dataset and labeling strategy described in Section~\ref{sec2.3:sample_label}, the evaluation is structured around two tasks.

Specifically, we first conduct a binary prediction task to assess the model’s ability to distinguish flaring ($\geq$ C-class) from no-flare samples, and subsequently perform a multi-class prediction task to evaluate its capability in discriminating specific flare classes among C, M, and X-class flares.

\subsection{Evaluation Metrics}
\begin{table}
	\centering
	\caption{Evaluation metrics of the model.}
	\label{tab05:metrics}
	\renewcommand{\arraystretch}{1.8}
	\begin{tabular}{lc}
		\toprule
		Metric & Definition \\
		\midrule  
		Accuracy & $ \dfrac{\mathrm{TP} + \mathrm{TN}}{\mathrm{TP} + \mathrm{TN} + \mathrm{FP} + \mathrm{FN}} $ \\
		Precision & $ \dfrac{\mathrm{TP}}{\mathrm{TP} + \mathrm{FP}} $ \\
		Recall & $ \dfrac{\mathrm{TP}}{\mathrm{TP} + \mathrm{FN}} $ \\
		FAR & $ \dfrac{\mathrm{FP}}{\mathrm{TP} + \mathrm{FP}} $ \\
		HSS & $ \dfrac{2(\mathrm{TP} \times \mathrm{TN} - \mathrm{FN} \times \mathrm{FP})}{(\mathrm{TP} + \mathrm{FN})(\mathrm{FN} + \mathrm{TN}) + (\mathrm{TP} + \mathrm{FP})(\mathrm{FP} + \mathrm{TN})} $ \\
		TSS & $ \dfrac{\mathrm{TP}}{\mathrm{TP} + \mathrm{FN}} - \dfrac{\mathrm{FP}}{\mathrm{FP} + \mathrm{TN}} $ \\
		F1 Score & $ 2 \times \dfrac{ \mathrm{Precision} \times \mathrm{Recall} }{ \mathrm{Precision} + \mathrm{Recall} } $ \\
		Frequency Bias & $ \dfrac{\mathrm{TP} + \mathrm{FP}}{\mathrm{TP} + \mathrm{FN}} $ \\
		\bottomrule 
	\end{tabular}	
\end{table}

To quantitatively evaluate the model’s performance, key metrics are computed based on the four fundamental components derived from the confusion matrix: true positives (TP), where positive samples are correctly predicted as positive; true negatives (TN), where negative samples are correctly predicted as negative; false positives (FP), where negative samples are incorrectly predicted as positive; and false negatives (FN), where positive samples are incorrectly predicted as negative. These components form the foundation for computing key performance metrics, including accuracy, precision, recall, False Alarm Rate (FAR), Heidke Skill Score (HSS), True Skill Statistic (TSS), Frequency Bias (FB), and F1-score, as defined in Table~\ref{tab05:metrics}.

In solar flare prediction tasks, the severe class imbalance often results in uneven sample distributions, which in turn affects the effectiveness of conventional evaluation metrics such as accuracy and precision \citep[e.g.,][]{bloomfield2012toward, bobra2015solar, li2024prediction}.
To ensure a more reliable evaluation, this study adopts TSS, HSS, F1-score, and FB as the primary performance metrics. TSS, ranging from $-1$ to $1$, measures the ability to distinguish between positive and negative classes and is insensitive to class imbalance. FB indicates systematic over- or underforecasting, a necessary complement to the TSS metric~\citep{leka2019comparison}.
HSS quantifies the improvement over random predictions, with scores ranging from $-\infty$ to $1$. The F1-score, defined as the harmonic mean of precision and recall, ranges from 0 to 1 and is widely adopted for evaluating performance in imbalanced prediction tasks due to its balanced consideration of false positives and false negatives.

\subsection{Experimental Setup}
All experiments are implemented using the pytorch framework and executed on an NVIDIA RTX 4090 GPU. The input magnetograms are uniformly resized to $240 \times 240$ pixels. For model optimization, we employ the AdamW optimizer with an initial learning rate of $10^{-5}$, a weight decay of 0.05, and a batch size of 32. We adopt a Cosine Annealing scheduler with Linear Warmup, where the learning rate gradually decays to a minimum. To further mitigate overfitting, we apply Stochastic Depth regularization with a drop-path rate of 0.2, and consistently employ Focal Loss as the loss function across all tasks. The maximum training epochs are set to 100, during which both training and testing losses are recorded to monitor learning dynamics.

\subsection{Experimental Result}
\begin{figure*}
	\label{fig05:two_four_loss}
	\centering
	\begin{subfigure}{0.45\textwidth}  
		\includegraphics[width=\linewidth]{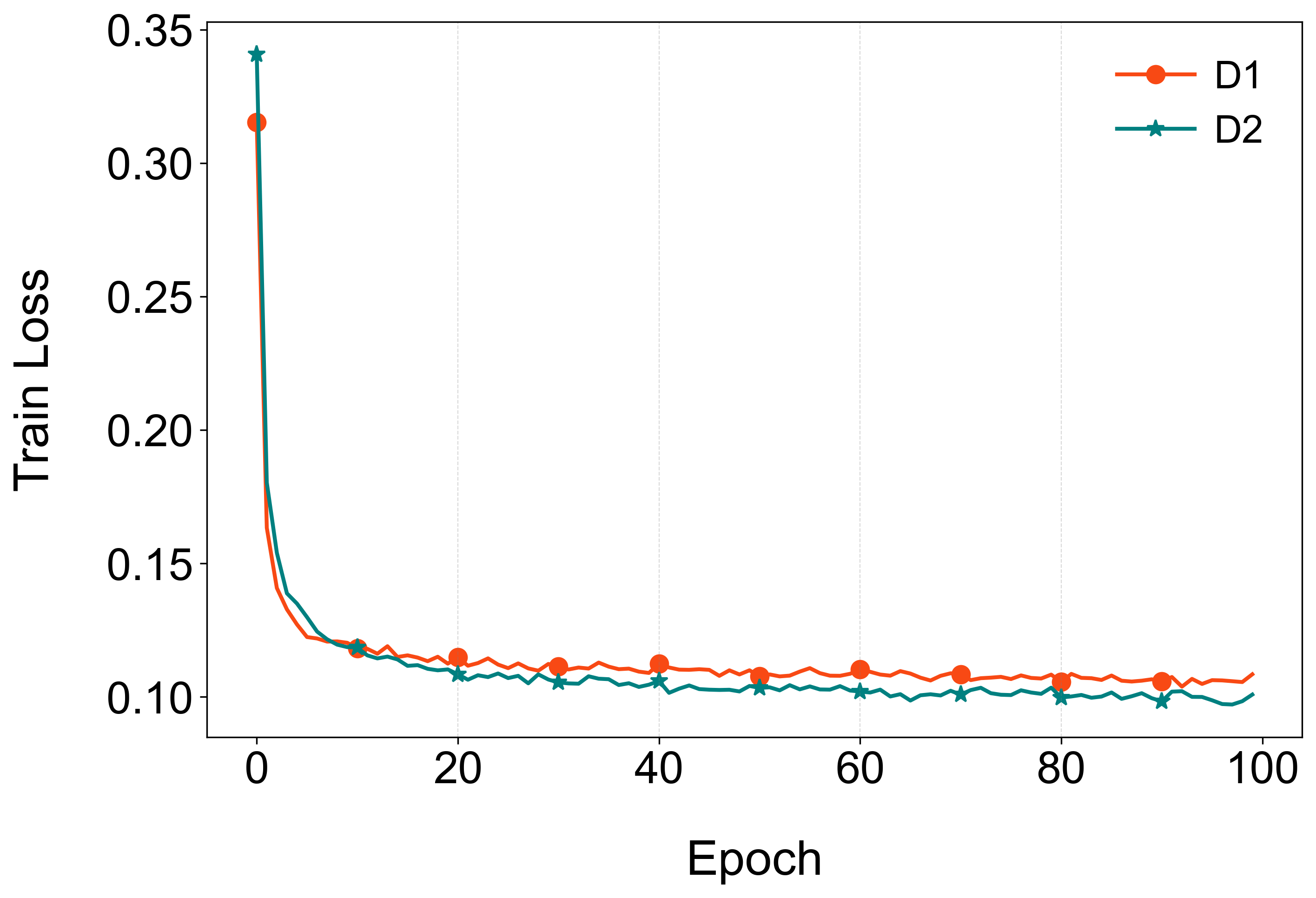}  
		\caption{}
		\label{fig05:two_train_loss}
	\end{subfigure}
	\hspace{0.05\textwidth}
	\begin{subfigure}{0.45\textwidth} 
		\includegraphics[width=\linewidth]{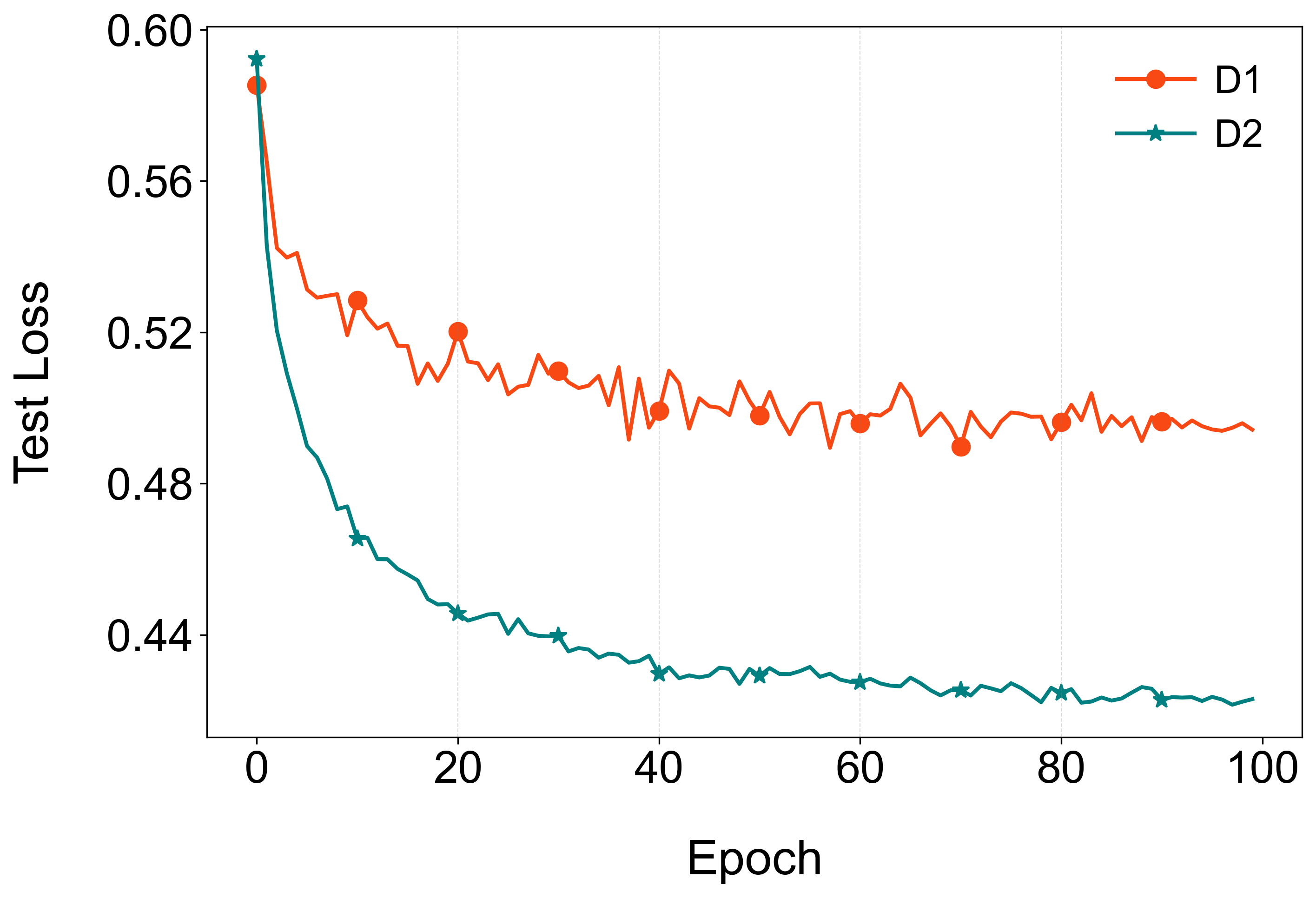}   
		\caption{}
		\label{fig05:two_test_loss}
	\end{subfigure}
	\begin{subfigure}{0.45\textwidth}  
		\includegraphics[width=\linewidth]{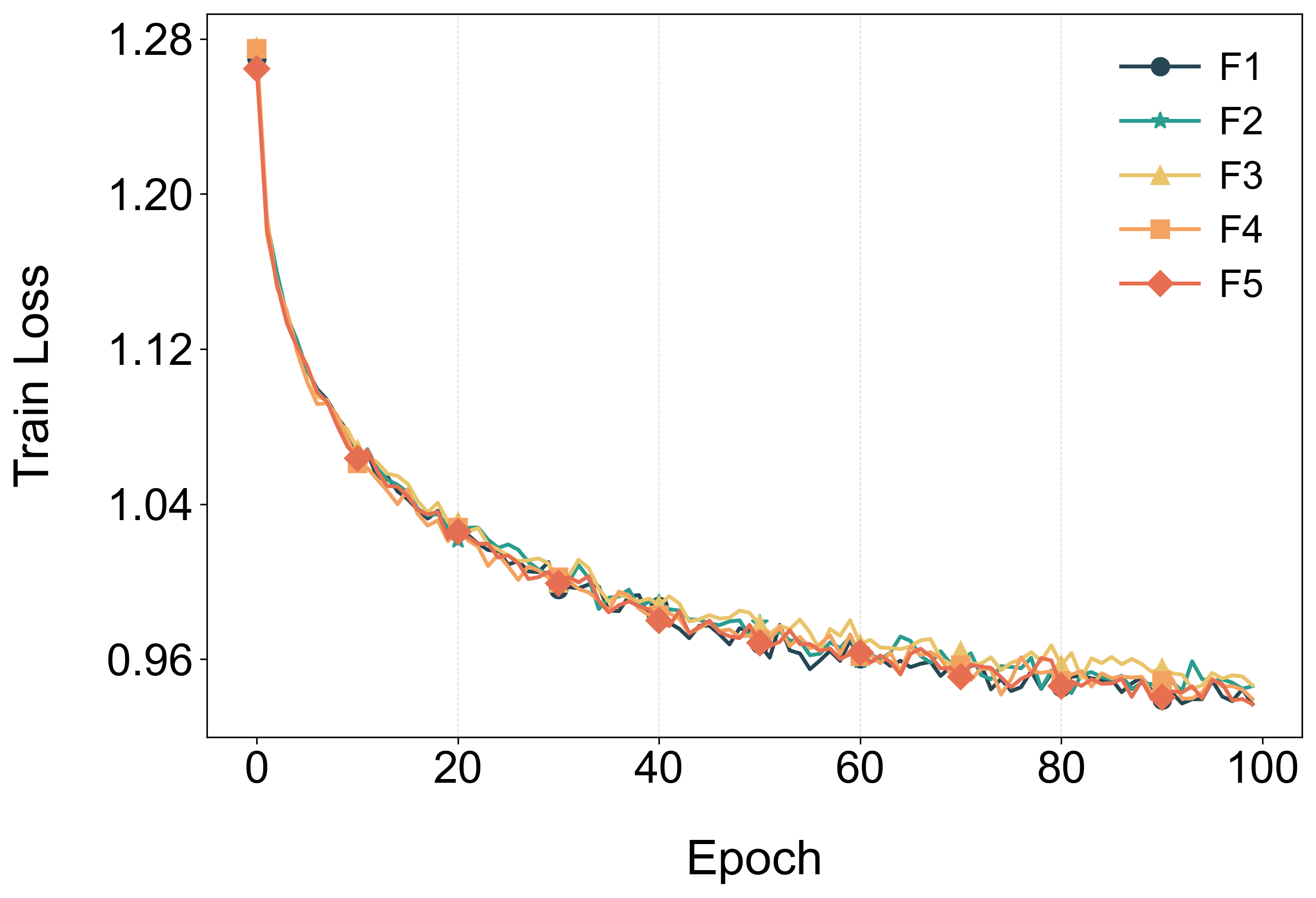} 
		\caption{}
		\label{fig05:four_train_loss}
	\end{subfigure}
	\hspace{0.05\textwidth}
	\begin{subfigure}{0.45\textwidth} 
		\includegraphics[width=\linewidth]{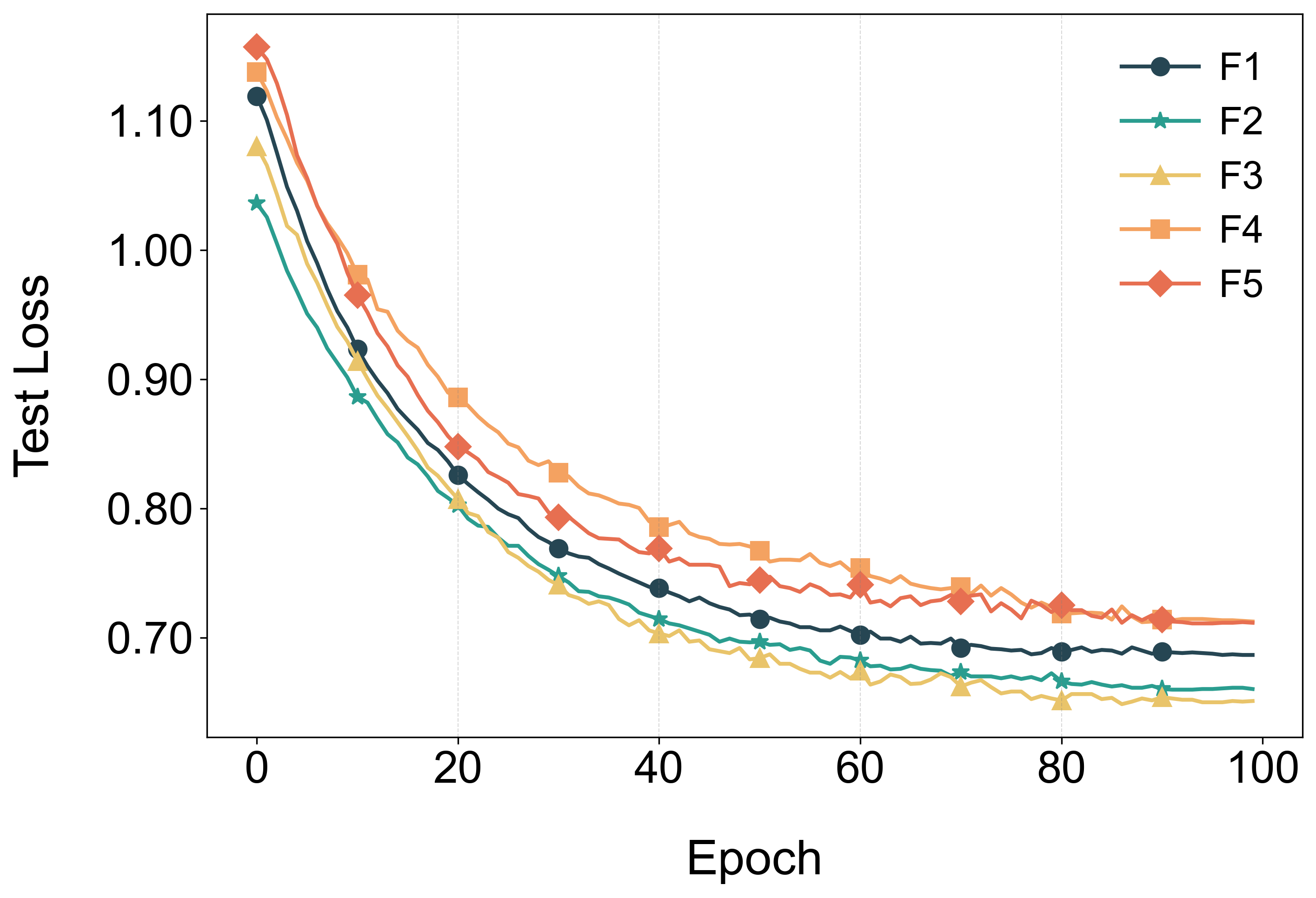}  
		\caption{}
		\label{fig05:four_test_loss}
	\end{subfigure}
	\caption{ The curves show the loss dynamics of our proposed model during training and testing. Curves of different colors represent the loss trajectories across epochs for different datasets. Subfigure (a) and (c) depicts the training loss variations for each dataset, while subfigure (b) and (d) shows the corresponding testing loss variations. }
\end{figure*}

To validate the effectiveness of the model and the stability of the training process, we first visualize the training and testing loss curves across different datasets. Figures~\ref{fig05:two_train_loss}–\ref{fig05:two_test_loss} illustrate the loss trajectories for the binary task of predicting $\geq$C-class flares, while Figures~\ref{fig05:four_train_loss}–\ref{fig05:four_test_loss} display the results for the multi-class discrimination of C, M, and X-class flares. As the number of epochs increases, the overall loss across all tasks decreases steadily, indicating that the model is learning effective representations.

For the binary prediction task, although the model achieves convergence on both datasets, the testing loss on the baseline dataset D1 is significantly higher than on D2, accompanied by noticeable fluctuations and a wider generalization gap, defined as the difference between training and testing losses. This indicates that on the imbalanced D1 dataset, the model predominantly learns the features of the majority no-flare samples, making it difficult to capture the key discriminative features of the minority flare class. In contrast, the significantly lower test loss observed on D2 implies that the model is able to focus more attention on the minority flare samples, thereby learning more distinguishable features rather than being dominated by the majority class.

In the multi-class task, Figures~\ref{fig05:four_train_loss}–\ref{fig05:four_test_loss} present the loss distributions across the five stratified group folds (F1-F5). The curves exhibit highly consistent trends with minimal variance between folds in both training and testing phases. This uniformity confirms that the model achieves stable convergence and robust generalization performance, ensuring that the evaluation is not biased by specific data splits.

\subsubsection{Binary prediction Results}
\begin{figure*}
	\centering
	\begin{subfigure}{0.45\textwidth}  
		\includegraphics[width=\linewidth]{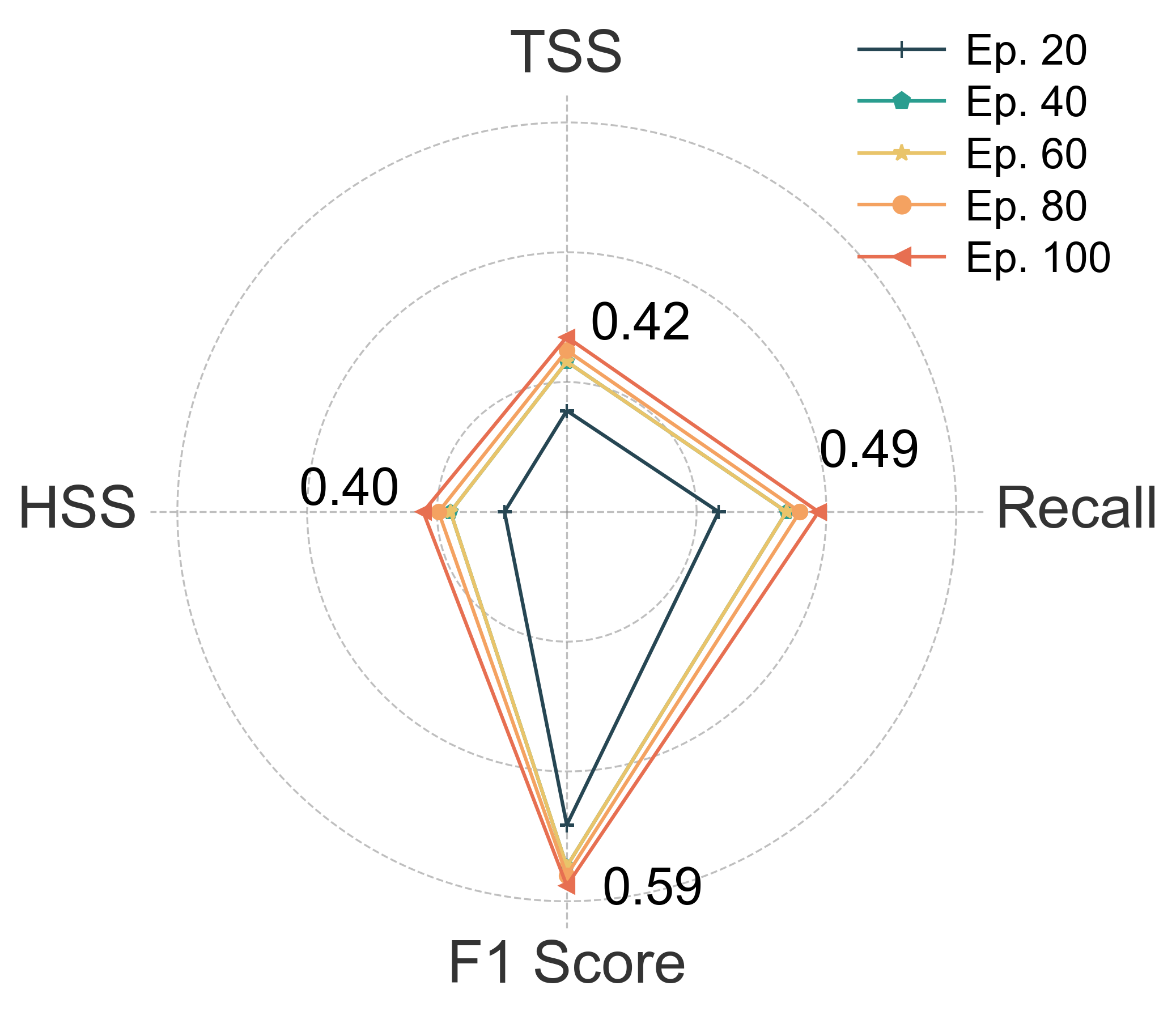} 
		\caption{}
		\label{fig06:D1}
	\end{subfigure}
	\hspace{0.05\textwidth}
	\begin{subfigure}{0.45\textwidth} 
		\includegraphics[width=\linewidth]{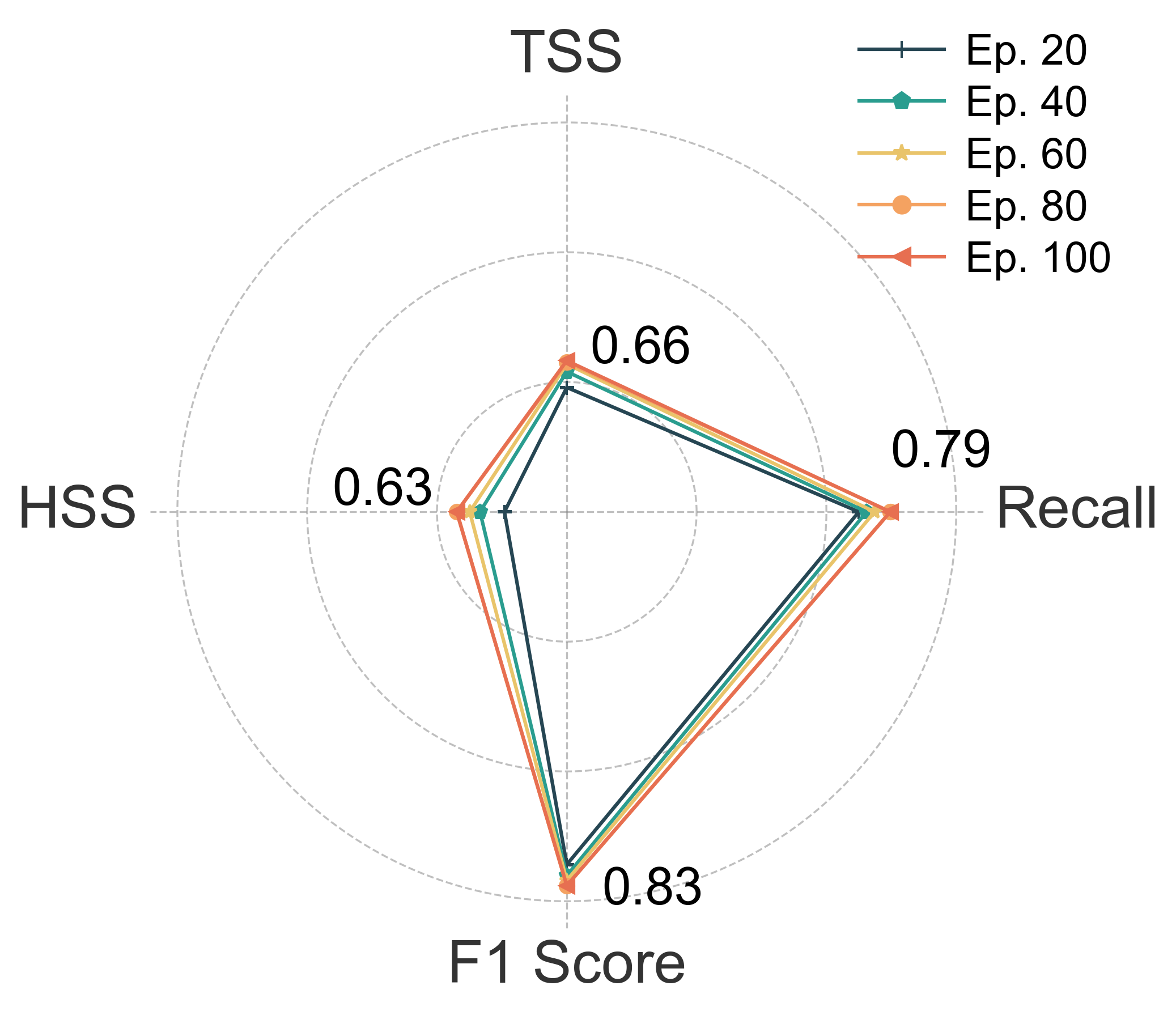}  
		\caption{}
		\label{fig06:D2}
	\end{subfigure}
	\caption{ The radar charts illustrate the evolution of recall, TSS, HSS,and F1 scores across training epochs on datasets D1 and D2. }
\end{figure*}

As shown in Figure~\ref{fig06:D1}, the model performance is first evaluated 
on D1. Although the model exhibits an upward trend throughout the training process, ultimately achieving an F1-score of $0.59$, skill metrics remain relatively limited, with TSS and HSS scores of only $0.42$ and $0.40$, respectively. Figure~\ref{fig06:D2} presents the evaluation results on the balanced D2. Compared with the results on D1, the discrepancy between skill scores and basic metrics narrows significantly. Specifically, recall rises to $0.79$, and both TSS and HSS exceed $0.63$, closely aligning with the high F1-score of $0.83$. 
The comprehensive improvement across key metrics indicates that the model has successfully captured the intrinsic characteristics of the minority class.

\begin{table} 
	\centering
	\caption{The flare prediction results of our model for $ \geqslant $C-class and a comparison with previous studies (within 24 hours).}
	\label{tab06:two_metrics} 
	\renewcommand{\arraystretch}{1.0}
	\setlength{\tabcolsep}{3.8mm}{
		\begin{tabular}{cccc} 
			\toprule
			Metric& \quad      Model&\qquad  $ \geqslant $C-class  \\
			\hline	
			\multirow{4}{*}{Recall}&\quad This work & \qquad 0.789 \\
			&\quad \cite{tang2021solar} & \qquad 0.592\\
			&\quad \cite{huang2018deep} & \qquad 0.726 \\
			&\quad \cite{nishizuka2018deep} & \qquad 0.809 \\
			\hline	
			\multirow{4}{*}{Precision}&This work & \qquad 0.868 \\
			&\quad \cite{tang2021solar} & \qquad 0.662 \\
			&\quad \cite{huang2018deep} & \qquad 0.352 \\
			&\quad \cite{nishizuka2018deep} &\qquad 0.529 \\
			\hline
			\multirow{4}{*}{Accuracy}&This work& \qquad 0.829 \\
			&\quad \cite{tang2021solar} & \qquad 0.820  \\
			&\quad \cite{huang2018deep} & \qquad 0.756\\
			&\quad \cite{nishizuka2018deep} & \qquad 0.822 \\
			\hline
			\multirow{4}{*}{HSS}&This work & \qquad 0.658  \\
			&\quad \cite{tang2021solar} & \: \textendash  \\
			&\quad \cite{huang2018deep} & \qquad 0.339 \\ 
			&\quad \cite{nishizuka2018deep}& \qquad 0.528\\
			\hline
			\multirow{4}{*}{TSS}&This work & \qquad 0.661  \\
			&\quad \cite{tang2021solar} & \qquad 0.535& \\
			&\quad \cite{huang2018deep} & \qquad 0.487 \\
			&\quad \cite{nishizuka2018deep}& \qquad 0.634\\
			\hline
			\multirow{4}{*}{F1 score}&This work & \qquad 0.827 \\
			&\quad \cite{tang2021solar} & \qquad 0.624 \\
			&\quad \cite{huang2018deep} & \qquad 0.475 \\
			&\quad \cite{nishizuka2018deep}& \: \textendash \\
			\hline
			\multirow{4}{*}{FB}&This work & \qquad 0.909 \\
			&\quad \cite{tang2021solar} & \: \textendash \\
			&\quad \cite{huang2018deep} & \: \textendash \\
			&\quad \cite{nishizuka2018deep}& \: \textendash \\
			\bottomrule
			
		\end{tabular}
	}
\end{table}
Based on these findings, Table~\ref{tab06:two_metrics} presents a quantitative comparison between our model and representative existing studies under unified prediction conditions. The results indicate that our model achieves superior performance across multiple key evaluation metrics. With TSS, HSS, and F1-score reaching $0.661$, $0.658$, and $0.827$, respectively, our model significantly outperforms the corresponding results reported by \cite{tang2021solar} and \cite{huang2018deep}. Although \citet{nishizuka2018deep} reported a recall of $0.809$, which is slightly higher than our $0.789$, this sensitivity is achieved at the cost of precision, which drops to $0.529$, indicating a tendency towards over-forecasting and a high false alarm rate. In contrast, our model achieves a high precision of $0.868$, demonstrating its ability to effectively suppress false alarms while maintaining high sensitivity. This robust balance between recall and precision is further reflected in the superior F1-score of $0.827$, which markedly exceeds those of the compared methods.

Furthermore, our model exhibits a FB of $0.909$, which is close to the ideal value of $1.0$. This result indicates that the model has no significant systematic bias in its overall prediction tendency, and its forecasted frequency aligns closely with actual observations. Overall, the proposed method demonstrates a more balanced and robust comprehensive performance across both basic and skill-based metrics.

\subsubsection{Multi-Class prediction Results}

\begin{figure*} 
	\centering
	\subfloat[]{
		\includegraphics[width=0.33\linewidth]{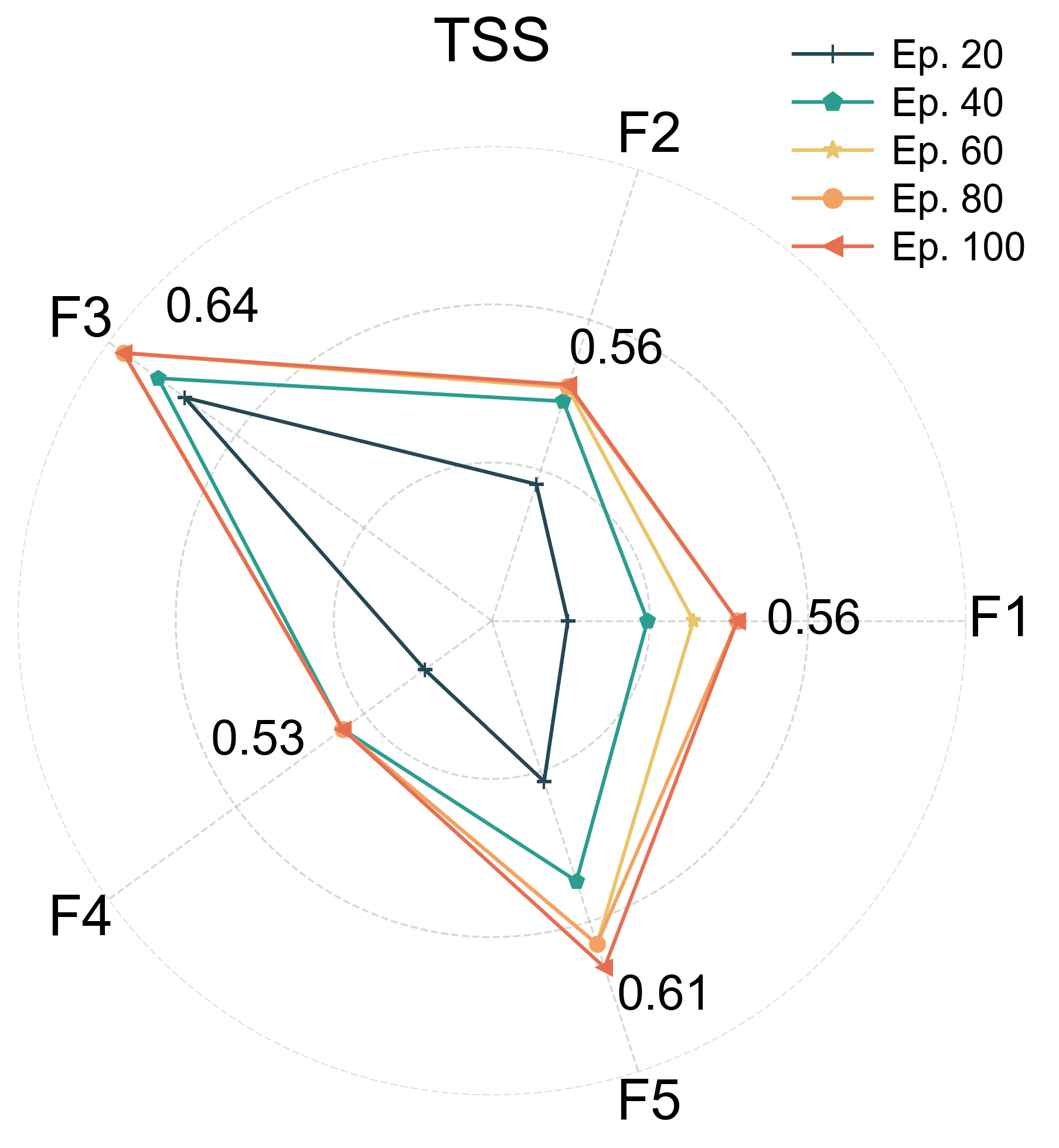}
		\label{C_class_tss}}
	\subfloat[]{
		\includegraphics[width=0.33\linewidth]{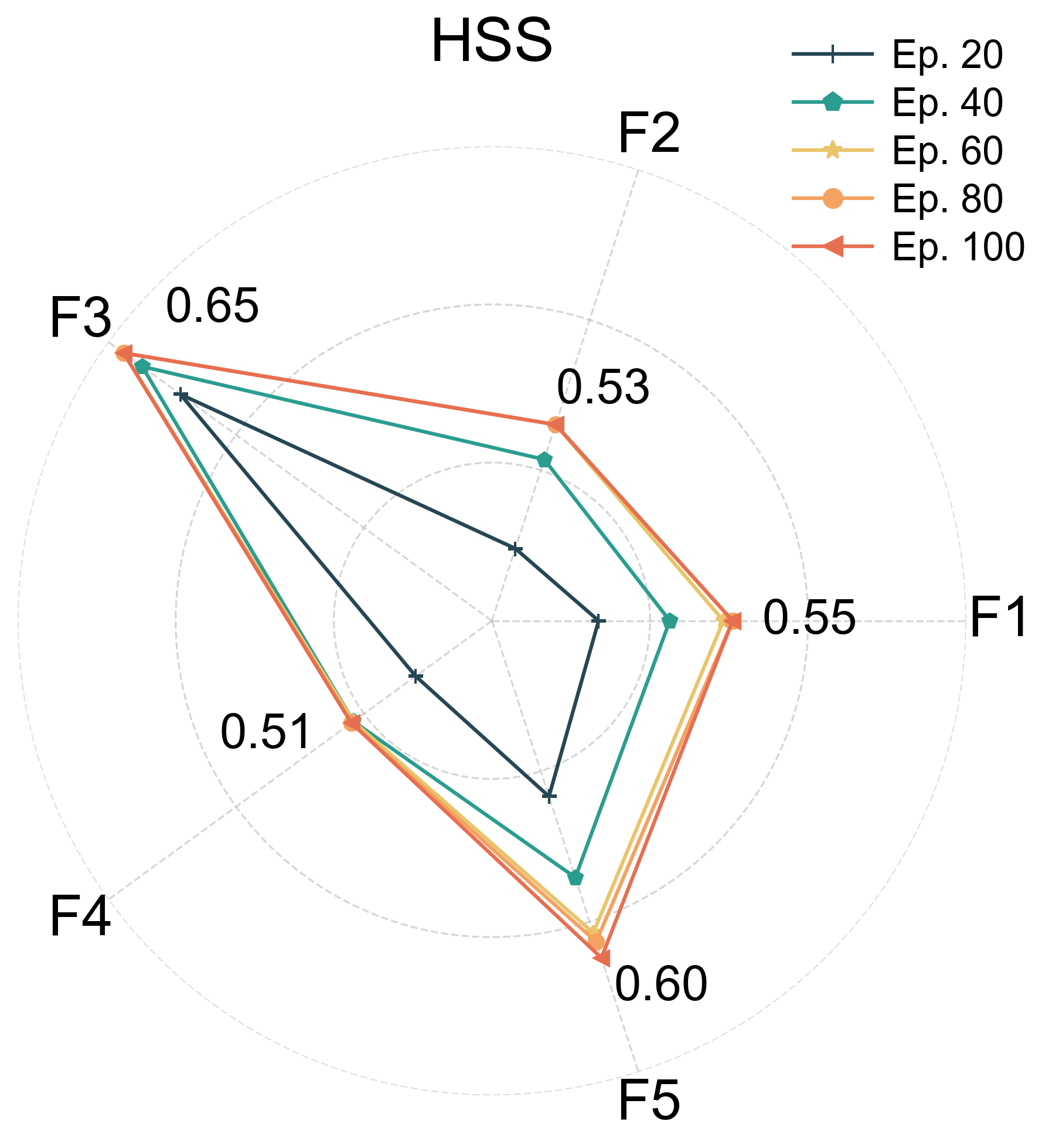}
		\label{C_class_hss}}
	\subfloat[]{
		\includegraphics[width=0.33\linewidth]{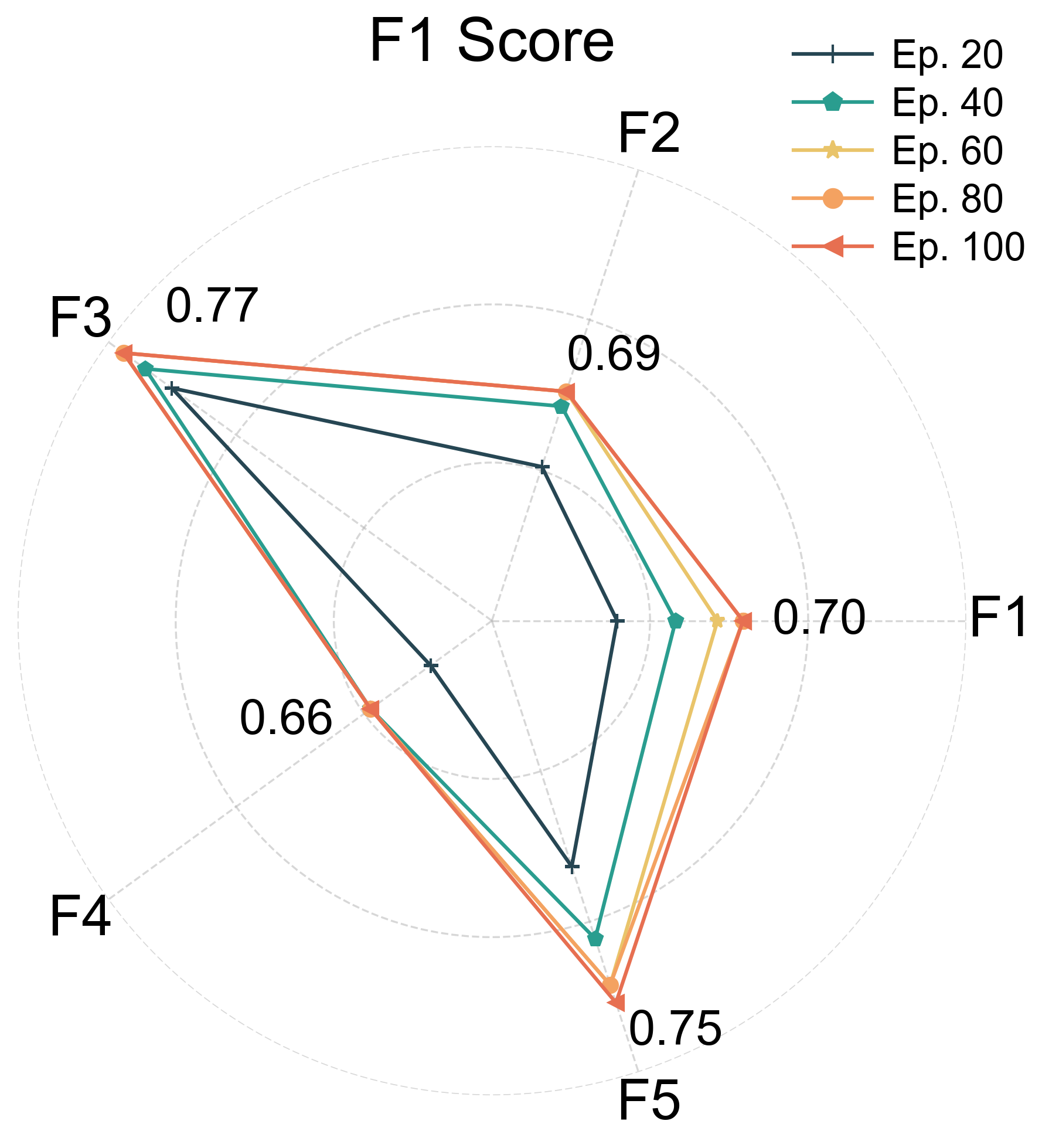}
		\label{C_class_f1}}\\
	
	\subfloat[]{
		\includegraphics[width=0.33\linewidth]{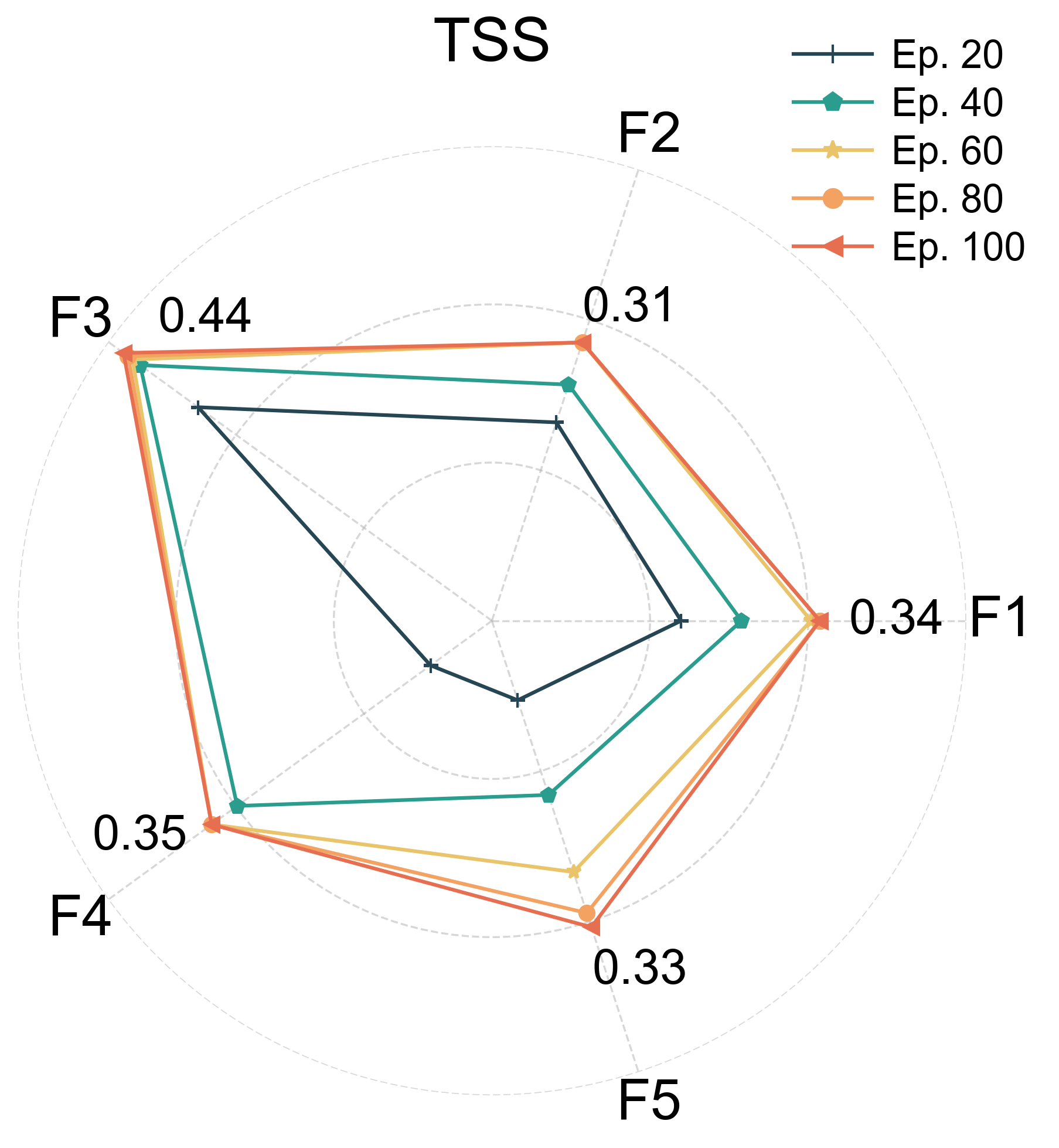}
		\label{M_class_tss}}  	
	\subfloat[]{
		\includegraphics[width=0.33\linewidth]{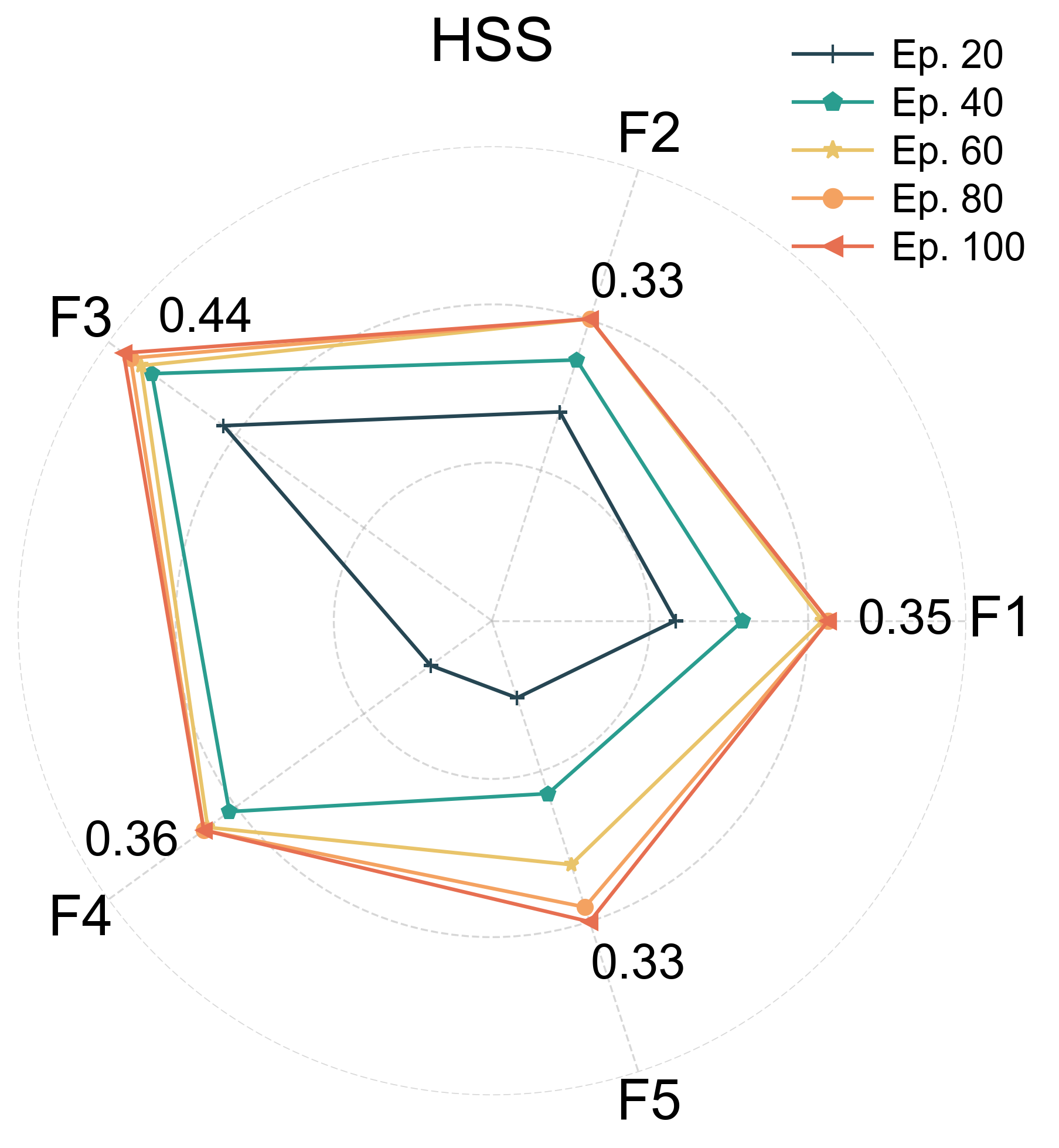}		
		\label{M_class_hss}}
	\subfloat[]{
		\includegraphics[width=0.33\linewidth]{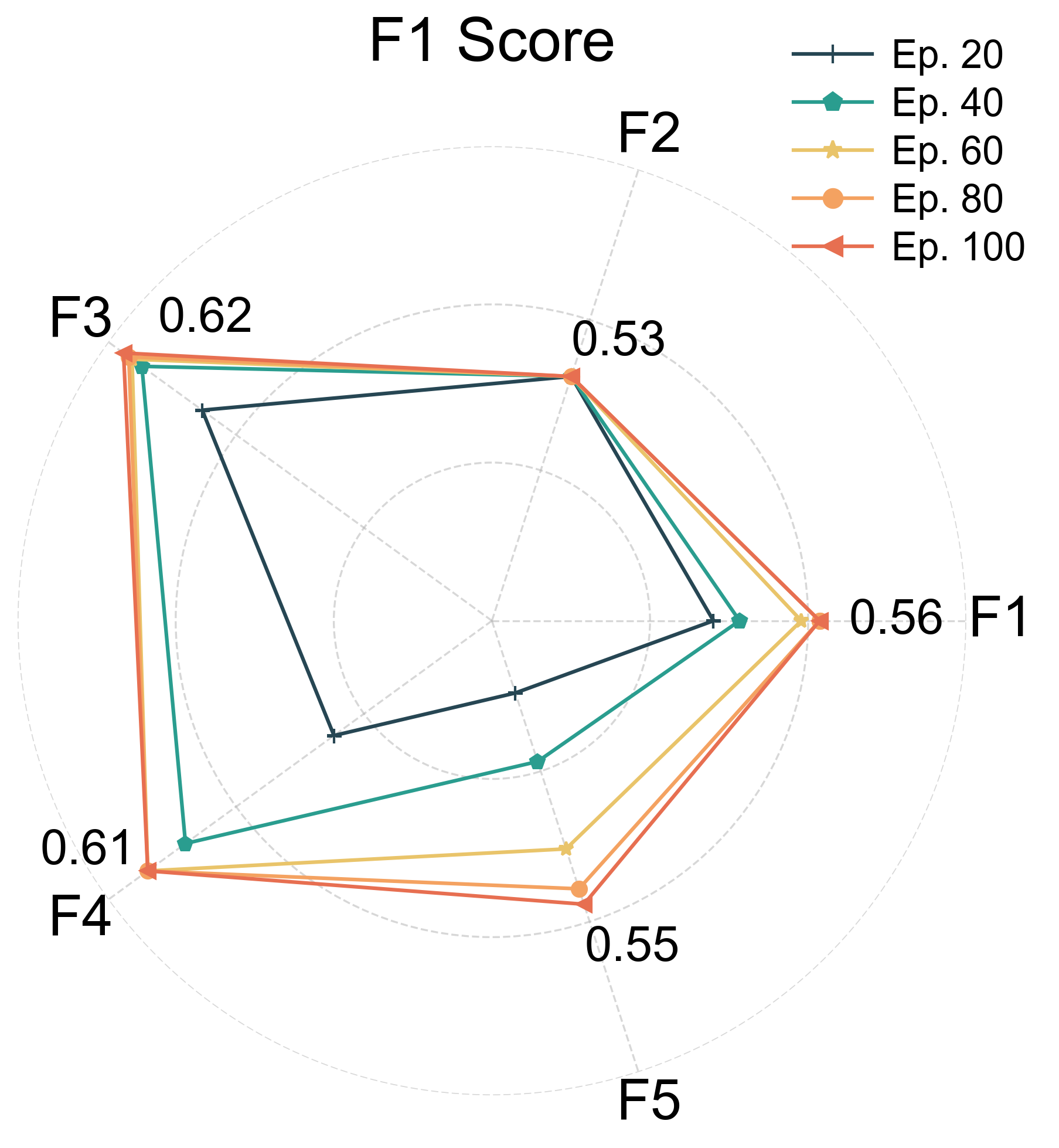} 
		\label{M_class_f1}}\\
	
	\subfloat[]{
		\includegraphics[width=0.33\linewidth]{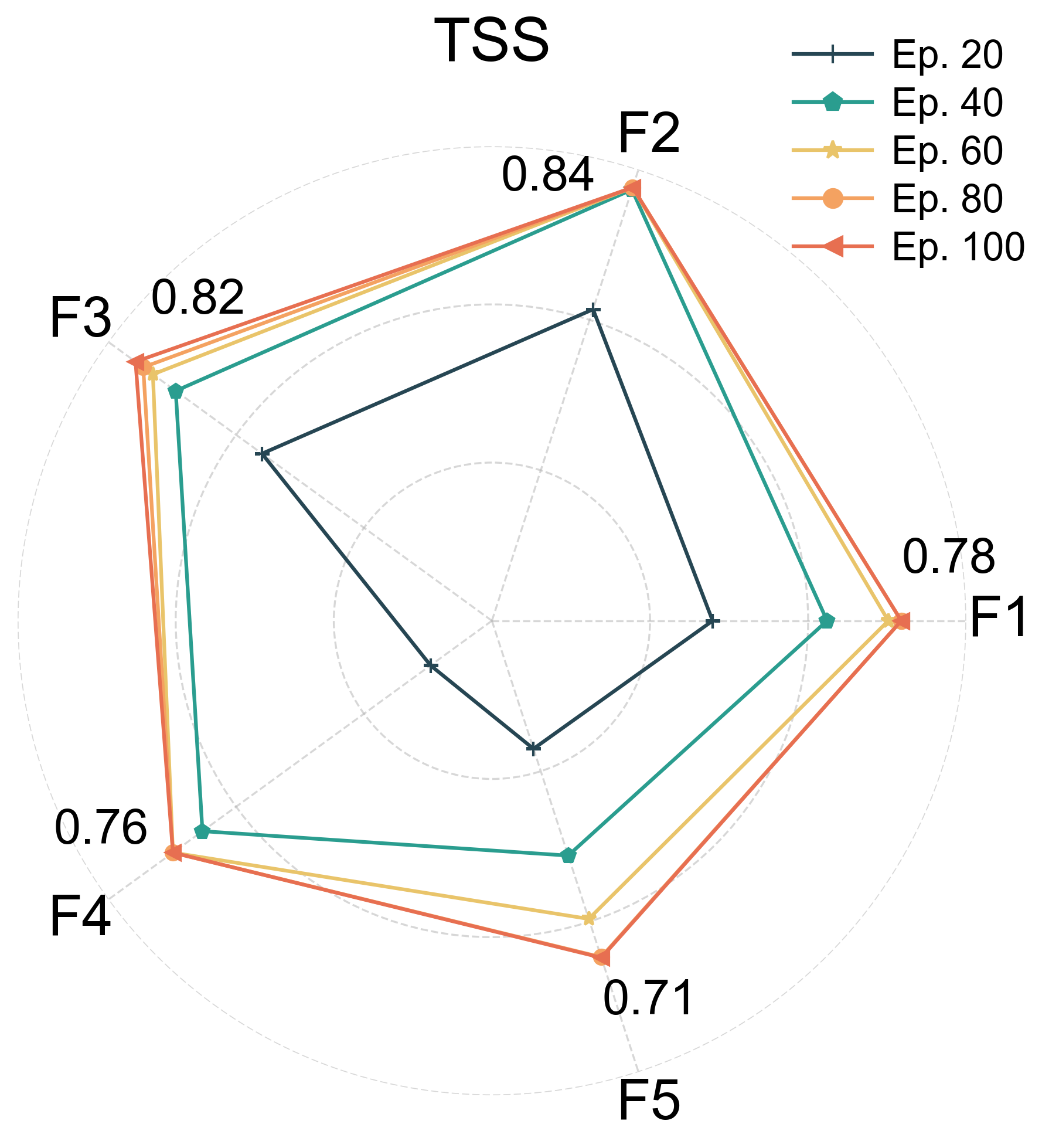}
		\label{X_class_tss}}
	\subfloat[]{
		\includegraphics[width=0.33\linewidth]{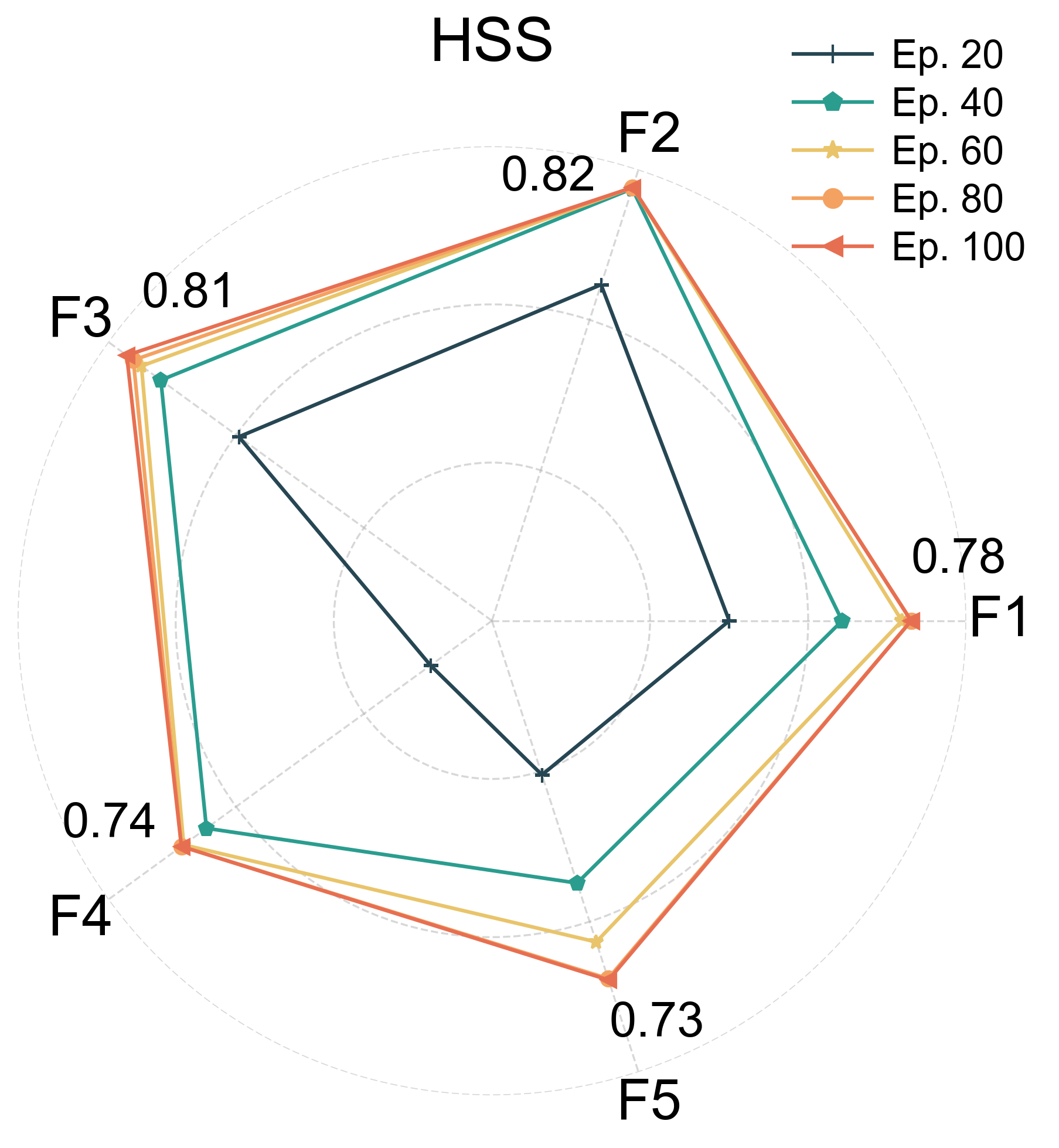}
		\label{X_class_hss}}
	\subfloat[]{
		\includegraphics[width=0.33\linewidth]{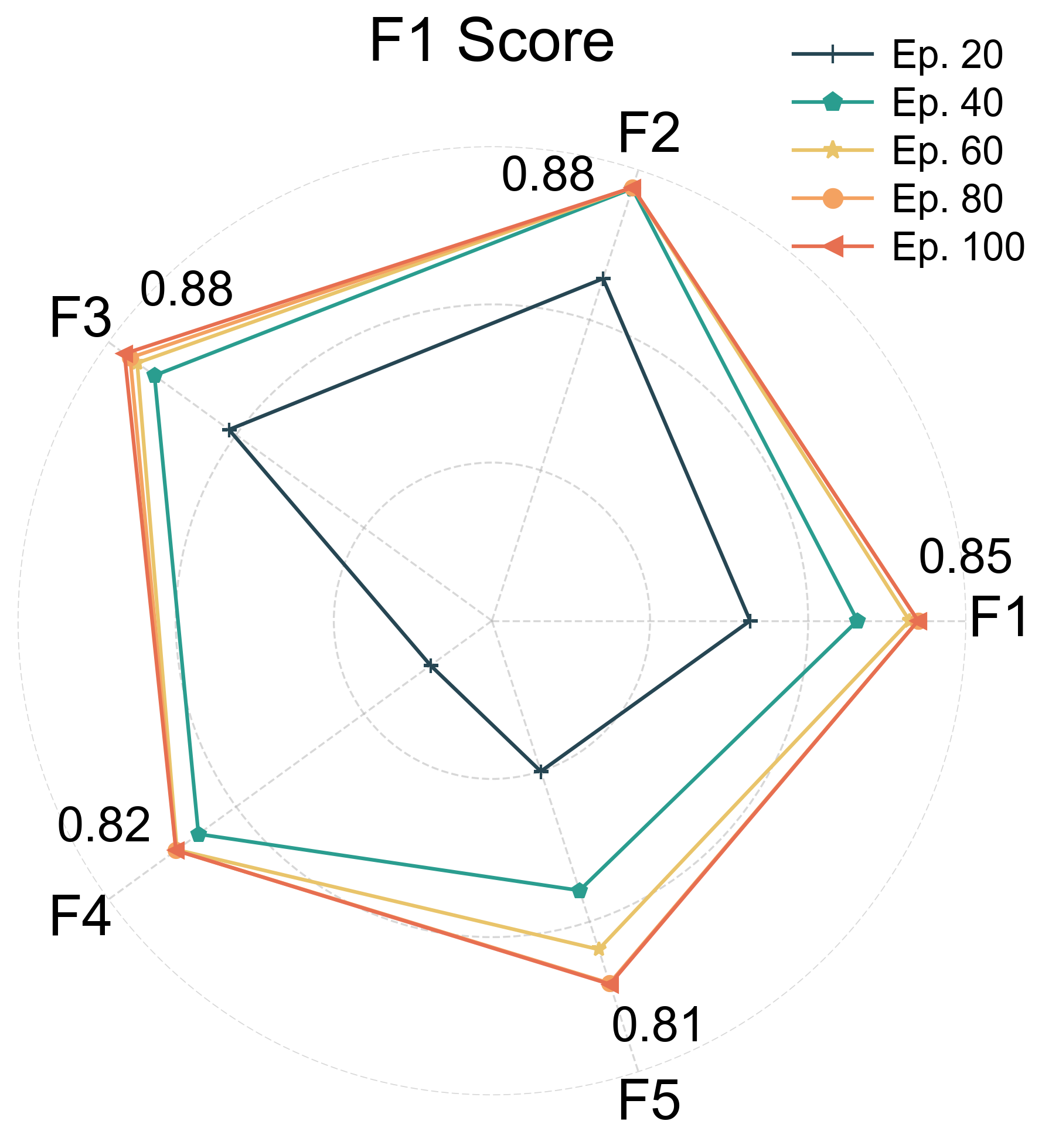}
		\label{X_class_f1}}
	\qquad 
	\caption{ The radar charts illustrate the evolution of TSS, HSS, and F1 scores for C, M, and X-class flares across training epochs under datasets F1–F5. The results are arranged vertically in the figure, with C-class at the top, M-class in the middle, and X-class at the bottom. }
	\label{fig07:radar_mertics}	
\end{figure*}

We analyze the performance of the model on five datasets (F1–F5). Figure~\ref{fig07:radar_mertics} shows the trends of key performance metrics for each flare class across training epochs. All metrics across F1-F5 show consistent improvement with increasing epochs, suggesting the model’s robust capacity for continual learning.

Most notably, for X-class flares, as illustrated in Figures~\ref{X_class_tss}–\ref{X_class_f1}, the model demonstrates superior and highly consistent performance. The metrics consistently fall within a high range, with TSS and HSS exceeding $0.80$ and the F1-score peaking at approximately $0.88$. Such outstanding performance validates the effectiveness of the multimodal fusion approach in extracting discriminative features from X-class magnetograms and their associated magnetic parameters. This may be attributed to the inherently distinctive and easily separable nature of X-class samples. Despite their scarcity, these samples exhibit pronounced differences from other flare types, thereby enabling more effective representation learning by the model.

For C-class flares, as illustrated in Figures \ref{C_class_tss}–\ref{C_class_f1}, the model maintains a solid baseline performance. The F1-score reaches up to $0.77$, while TSS and HSS generally stabilize in the range of $0.51$ to $0.65$. Notably, the F3 and F5 subsets consistently yield the best performance across all three indicators, indicating that their sample feature distributions are more favorable for modeling C-class flares. Although these scores are lower than those of the X-class, they represent a reasonable balance between precision and recall in identifying C-class events.

However, the primary challenge lies in M-class prediction. As shown in Figures~\ref{M_class_tss}–\ref{M_class_f1}, the performance metrics exhibit noticeable fluctuations compared to the C and X-classes, with HSS and TSS remaining in relatively low ranges and limited improvement in F1-score. This U-shaped performance profile suggests that M-class flares likely possess overlapping features with both weaker (C-class) and stronger (X-class) flares, making the decision boundaries more difficult to resolve. Nevertheless, the upward trend in metrics throughout the training epochs indicates that the model is gradually learning to mitigate this ambiguity over time.

\begin{table*} 
	\centering
	\caption{ The Multiclass flare prediction results of our model and a comparison with previous studies (within 24 hours).}
	\label{tab07:four_metrics}
	\renewcommand{\arraystretch}{1.0}
	\setlength{\tabcolsep}{4.7mm}{
		\begin{tabular}{ccccccc}
			\toprule
			Metric& Model& C & M& X \\
			\hline	
			\multirow{4}{*}{Recall}&This work  &0.742 ± 0.030&0.551 ± 0.058& 0.857 ± 0.059 \\
			&\cite{zheng2021hybrid}&0.644 $\pm$  0.083&0.518 $\pm$ 0.272& 0.523 $ \pm $ 0.398 \\
			&\cite{zheng2019solar} &0.671 $\pm$ 0.059&0.617 $\pm$ 0.148& 0.594 ± 0.394\\
			& \cite{liu2017predicting}&0.526 $\pm$ 0.050&0.671 $\pm$ 0.037& 0.297 ± 0.039 \\
			\hline	
			\multirow{4}{*}{Precision}&This work&0.686 ± 0.066&0.608 ± 0.048& 0.840 ± 0.035 \\
			&\cite{zheng2021hybrid}   &0.638 $\pm$ 0.041&0.659 $\pm$ 0.060& 0.502 $\pm$ 0.346 \\
			&\cite{zheng2019solar}    &0.670 $\pm$ 0.079&0.699 $\pm$ 0.087&0.562 ± 0.383\\
			& \cite{liu2017predicting}&0.563 $\pm$ 0.054&0.656 $\pm$ 0.036& 0.745 $\pm$ 0.152\\
			\hline
			\multirow{4}{*}{Accuracy} &This work&0.806 ± 0.023&0.714 ± 0.025& 0.902 ± 0.015 \\
			&\cite{zheng2021hybrid}   &0.791 $\pm$ 0.039&0.830 $\pm$ 0.031& 0.893 $\pm$ 0.065 \\
			&\cite{zheng2019solar}    & 0.812 $\pm$  0.029&0.849 $\pm$  0.034&0.933 $\pm$  0.041\\
			& \cite{liu2017predicting}&0.712 $\pm$ 0.026&0.778 $\pm$ 0.019& 0.957 $\pm$ 0.005 \\
			\hline
			\multirow{4}{*}{FAR}&This work &0.314 ± 0.066&0.392 ± 0.048& 0.160 ± 0.035 \\
			&\cite{zheng2021hybrid} &0.361 $\pm$ 0.041&0.340 $\pm$ 0.060& 0.297 $\pm$ 0.281 \\
			&\cite{zheng2019solar} &0.330 $\pm$ 0.079&0.301 $\pm$ 0.087& 0.138 $\pm$ 0.140 \\
			& \cite{liu2017predicting}&0.437 $\pm$ 0.016&0.344 $\pm$ 0.020& 0.255 $\pm$ 0.126 \\
			\hline
			\multirow{4}{*}{HSS}&This work &0.566 ± 0.056&0.361 ± 0.047 & 0.775 ± 0.041 \\
			&\cite{zheng2021hybrid} &0.488 $\pm$ 0.047&0.444 $\pm$ 0.194& 0.419 $\pm$ 0.309 \\
			&\cite{zheng2019solar} &0.535 $\pm$ 0.061&0.551 $\pm$ 0.120& 0.539 ± 0.366\\
			& \cite{liu2017predicting}&0.334 $\pm$ 0.028&0.497 $\pm$ 0.031& 0.406 $\pm$ 0.014\\
			\hline
			\multirow{4}{*}{TSS}&This work&0.579 ± 0.045&0.356 ± 0.051& 0.780 ± 0.051 \\
			&\cite{zheng2021hybrid} &0.489 $\pm$ 0.049&0.432 $\pm$ 0.222& 0.436 $\pm$ 0.330 \\
			&\cite{zheng2019solar} &0.538 $\pm$ 0.059&0.534 $\pm$ 0.137& 0.552 $\pm$ 0.370\\
			&\cite{liu2017predicting}&0.328 $\pm$ 0.050&0.500 $\pm$ 0.037& 0.291 $\pm$ 0.039 \\
			\hline
			
			\multirow{4}{*}{FB}&This work &1.087 ± 0.091&0.912 ± 0.131& 1.022 ± 0.088 \\
			&\cite{zheng2021hybrid} &\: \textendash&\: \textendash& \: \textendash \\
			&\cite{zheng2019solar} &\: \textendash&\: \textendash& \: \textendash \\
			& \cite{liu2017predicting}&\: \textendash&\: \textendash& \: \textendash \\
			\hline
			\multirow{4}{*}{F1 score}&This work &0.712 ± 0.043&0.576 ± 0.037& 0.847 ± 0.032 \\
			&\cite{zheng2021hybrid} &0.489 $\pm$ 0.049&0.432 $\pm$ 0.222& 0.436 $\pm$ 0.330 \\
			&\cite{zheng2019solar} &\: \textendash&\: \textendash& \: \textendash \\
			& \cite{liu2017predicting}&\: \textendash&\: \textendash& \: \textendash \\
			\bottomrule
		\end{tabular}
	}
\end{table*}

To evaluate the model's predictive performance across multiple classes (C, M, and X), our average results are compared with those fromm prior studies that conducted predictions (see Table \ref{tab07:four_metrics}).

For X-class flares, the proposed method demonstrates superior performance, significantly outperforming existing approaches \cite[e.g.,][]{liu2017predicting, zheng2019solar,zheng2021hybrid} in terms of HSS ($0.775$) and TSS ($0.780$). Notably, the FAR is remarkably low at $0.160$, indicating a minimal false positive rate. Moreover, the frequency bias of $1.022 \pm 0.088$ suggests that the model exhibits negligible systematic bias, effectively avoiding the tendency towards significant over- or under-forecasting.
Additionally, the F1-score reaches $0.847$, reflecting a well-balanced performance between precision and recall. These results highlight the model's high sensitivity and robust predictive capability for intense flare events, underscoring its potential utility in severe space weather forecasting.

For C-class flares, the model achieves robust results, with an HSS of $0.566$ and a TSS of $0.579$—both significantly higher than those reported by \cite{zheng2021hybrid,zheng2019solar,liu2017predicting}. The F1-score reaches $0.712 \pm 0.043$, notably outperforming the $0.489$ reported by \cite{zheng2021hybrid}. Furthermore, the FAR is reduced to $0.314 \pm 0.066$, which is markedly lower than the values reported by \cite{zheng2019solar} ($0.330 \pm 0.079$), \cite{liu2017predicting} ($0.437 \pm 0.016$), and \cite{zheng2021hybrid} ($0.361 \pm 0.041$). These metrics indicate that the model not only effectively identifies C-class flares but also substantially suppresses false alarms, demonstrating its practical applicability. While M-class performance shows a slight compromise compared to previous works, this trade-off is strategically acceptable. The slight dip in M-class performance has enabled the model to prioritize and maximize the detection capabilities for the most destructive X-class events, which is the primary objective of operational space weather forecasting.

In conclusion, the proposed dual-branch multimodal fusion model demonstrates exceptional robustness and discriminative ability, particularly in identifying X-class flares. Compared with existing approaches \citep{zheng2021hybrid, zheng2019solar, liu2017predicting}, our model achieves consistently superior performance across multiple metrics for X-class and C-class events, while effectively reducing the false alarm rate. Although the model exhibits slightly lower performance on certain metrics for M-class flares compared to some previous studies, its overall performance remains competitive. These findings are well aligned with the core research objective of focusing on major solar flare eruptions.

It should be noted that the current field lacks a standardized and publicly available test set, making it difficult to conduct fair and quantitative comparisons across different methods \cite[e.g.,][]{barnes2016comparison,tang2021solar}. 
Therefore, the results presented are intended to serve as a valuable point of reference rather than a definitive performance comparison, demonstrating the proposed method's potential for operational early warning applications.

\subsection{Ablation Studies}
To thoroughly verify the effectiveness of the model and investigate the specific contributions of key components and experimental settings, we design and implement four sets of rigorous ablation studies.

First, we establish a model relying solely on line-of-sight magnetograms (corresponding to the visual branch in Figure~\ref{fig02:2_CNN}) as the Magnetogram-only baseline. By comparing this baseline with the complete fusion model, we quantify the performance gains introduced by the physical parameter branch. Subsequently, we evaluate the interaction efficiency of different feature fusion strategies. Furthermore, addressing details of data preprocessing and model architecture, we analyze the effects of input image resolution and the distinct roles of Positional Encoding (PE) within the magnetogram and physical parameter branches. In conclusion, these experiments verify the effectiveness and robustness of the proposed dual-branch multimodal fusion model from multiple aspects.

\subsubsection{Multimodal Inputs}
\begin{table}
	\renewcommand{\arraystretch}{1.1}
	\setlength{\tabcolsep}{0pt}      
	\begin{threeparttable}
		\caption{Performance comparison among the multimodal fusion model, magnetogram-only baseline, and physical parameter-only baseline.}
		\label{tab08:Ab_fusion}
		\begin{tabular*}{\linewidth}{@{\extracolsep{\fill}}lccccccccc}
			\toprule
			Model & Mag. & Param. & PE$_{\text{Mag}}$ & PE$_{\text{Param}}$ & HSS & TSS & FB & FAR & ACC \\
			\midrule
			\textbf{Fusion} & $\cmark$ & $\cmark$ & $ \cmark $ &  $\xmark$  & 0.658 & 0.661 & 0.909 & 0.132 & 0.829 \\
			Mag-only& $\cmark$ & $ \xmark $ & $\cmark$ & $ \xmark $ & 0.619 & 0.646 & 0.919 & 0.143 & 0.822 \\		
			Param-only& $\xmark$ & $ \cmark $ & $\xmark$ & $ \xmark $ & 0.527 & 0.527 & 1.0002 & 0.228 & 0.764 \\		
			\bottomrule
		\end{tabular*} 
		\begin{tablenotes}
			\item
			Mag. and Param. denote the magnetogram and magnetic parameter modalities, respectively. PE$_{\text{Mag}}$ and PE$_{\text{Param}}$ indicate that positional encoding is applied only to the magnetogram branch and the physical parameter branch, respectively.
		\end{tablenotes}
	\end{threeparttable}
\end{table}
To validate the effectiveness of the proposed multimodal fusion strategy, we conduct a comprehensive performance comparison among the complete fusion model and the two single-modality baselines, namely the magnetogram-only model with the parameter branch removed and the magnetic parameter-only model using only the SHARP parameters as input. This experiment investigates whether integrating strong physical priors as a supplement to implicit visual features can provide critical complementary information to better capture flare precursor signatures.

The experimental results are presented in Table~\ref{tab08:Ab_fusion}. The fusion model achieves the best overall performance with a peak TSS of 0.661. While both single-modality baselines exhibit slightly better FB values, their independent predictive capabilities are compromised by elevated false alarm rates, which results in a decline in key metrics such as ACC, TSS, and HSS. Compared to the magnetogram-only baseline, the multimodal approach indicates that magnetic parameters, acting as highly condensed physical representations, provide stable global constraints for the model. This synergy between the two modalities effectively suppresses local noise interference, thereby markedly improving the accuracy and stability of flare predictions.

\subsubsection{Feature Fusion Strategies}
\begin{figure} 
	\centering
	\includegraphics[width=0.7\linewidth]{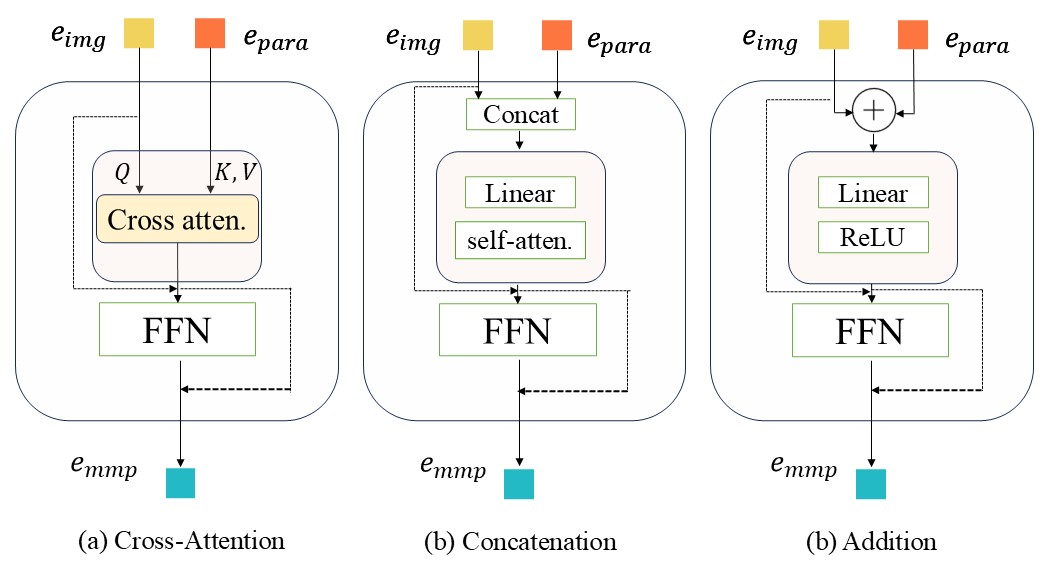}
	\caption{Comparison of multimodal interaction strategies between magnetograms and physical parameters.}
	\label{fig08:CA_add_concat}
\end{figure}
\begin{table}
	\renewcommand{\arraystretch}{1.1}
	\setlength{\tabcolsep}{0pt} 
	\begin{threeparttable}
		\caption{\makebox[\linewidth][l]{Performance comparison of different feature fusion strategies.}}
		\label{tab09:Ab_CA_add_Concat}
		\begin{tabular*}{\linewidth}{@{\extracolsep{\fill}}lccccccccc}
			\toprule
			Model & Mag. & Param. &  PE$_{\text{Mag}}$ & PE$_{\text{Param}}$ & HSS & TSS & FB & FAR & ACC \\
			\midrule
			\textbf{CA} & $\cmark$ & $\cmark$ & $ \cmark $ & $\xmark$ & 0.658 & 0.661 & 0.909 & 0.132 & 0.829 \\
			Concat. & $\cmark$ & $\cmark$ & $ \cmark $ & $\xmark$ & 0.635 & 0.653 & 0.947 & 0.150 & 0.826 \\
			Add.    & $\cmark$ & $\cmark$ & $ \cmark $ & $\xmark$ & 0.620 & 0.648 & 0.914 & 0.140 & 0.823 \\
			Inv-CA  & $\cmark$ & $\cmark$ & $ \cmark $ & $\xmark$ & 0.623 & 0.646 & 0.931 & 0.147 & 0.822 \\	
			\bottomrule
		\end{tabular*}
		\begin{tablenotes}
			\item Same as in Table \ref{tab08:Ab_fusion}. CA, Concat., and Add. denote the Cross-Attention, Concatenation, and direct Addition fusion strategies, respectively. Inv-CA denotes the inverse Cross-Attention strategy, where physical parameters serve as queries and magnetograms serve as keys/values.	
		\end{tablenotes}
	\end{threeparttable}	
\end{table}
Serving as a critical bridge between the magnetogram and physical parameter modalities, the multimodal interaction module plays a core role in enhancing the discriminability and feature alignment of cross-modal representations. As illustrated in Figure \ref{fig08:CA_add_concat}, we systematically explore various feature fusion strategies. Despite variations in their specific architectures, all strategies follow a unified design principle: using the magnetogram modality as the primary information stream and incorporating parameter information to enrich feature representation and achieve cross-modal coordination. The corresponding quantitative comparison are summarized in Table \ref{tab09:Ab_CA_add_Concat}.

We first compare the effectiveness of different fusion operations. The concatenation strategy achieves modality integration by feature concatenation and linear projection, followed by self-attention within the joint feature space to obtain fused representations. Although its TSS (0.653) is relatively close to that of the cross-attention fusion strategy (0.661), it exhibits a noticeably higher FAR, accompanied by a decline in overall accuracy. This suggests that simple linear concatenation has limitations in modeling cross-modal dependencies and may introduce additional false alarms. In contrast, the direct addition strategy learns fused features by combining linear mappings and non-linear transformations. However, the results indicate that this approach offers no distinct advantages across key metrics such as TSS, FAR, and ACC. Compared with the cross-attention strategy, it fails to yield benefits in either predictive accuracy or computational efficiency.

Furthermore, to investigate the impact of cross-modal interaction direction on performance, we compare two attention mechanism designs. In the standard cross-attention, we use the magnetogram as the query, while physical parameter features serve as keys and values, thereby embedding physical prior information into the spatial magnetogram representations. To validate the necessity of this design, we construct a variant model named Param-Q (Row 4), which conversely treats physical parameters as the query, attempting to integrate the visual information of magnetograms into parameter features. However, experimental results indicate a significant performance drop for Param-Q, even falling below the direct addition strategy. This result strongly suggests that, within the proposed model, using the magnetogram as the primary feature carrier is more conducive to preserving discriminative spatial structural features while embedding complementary magnetic parameter information.

Based on this comparative analysis, we adopt the cross-attention mechanism as the fusion strategy for the multimodal interaction module.

\subsubsection{Positional Encoding}
\begin{table}
	\renewcommand{\arraystretch}{1.1}
	\setlength{\tabcolsep}{0pt} 
	\begin{threeparttable}
		\caption{Performance comparison of the model with and without positional encoding on magnetogram and parameter branches.}
		\label{tab10:Ab_PE}
		\begin{tabular*}{\linewidth}{@{\extracolsep{\fill}}lccccccccc}
			\toprule
			Model & Mag. & Param. & PE$_{\text{Mag}}$ & PE$_{\text{Param}}$ & HSS & TSS & FB & FAR & ACC \\
			\midrule
			\textbf{Mag-PE} & $\cmark$ & $\cmark$ & $\cmark$ & $ \xmark $ & 0.658 & 0.661 & 0.909 & 0.132 & 0.8288 \\
			No-PE  & $\cmark$ & $\cmark$ & $ \xmark $ & $ \xmark $ & 0.645 & 0.658 & 0.964 & 0.153 & 0.8284 \\
			All-PE & $\cmark$ & $\cmark$ & $\cmark$ & $\cmark$ & 0.635 & 0.659 & 0.927 & 0.139 & 0.8285 \\
			\bottomrule
		\end{tabular*} 
		\begin{tablenotes}
			\item
			Same as in Table \ref{tab08:Ab_fusion}. Mag-PE denotes positional encoding only to the magnetogram branch; No-PE represents the setting without positional encoding, and All-PE applies positional encoding to all modalities.		
		\end{tablenotes}
	\end{threeparttable}	
\end{table}  
Positional information plays a crucial role in characterizing the spatial morphological evolution of active regions. To systematically analyze the role of positional encoding (PE) across different modalities, as shown in Table~\ref{tab10:Ab_PE}, we design and compare three PE configuration strategies, applying PE only to the magnetogram branch (Mag-PE), applying PE to neither branch (No-PE), and applying PE to both branches simultaneously (All-PE).

The experimental results demonstrate that the Mag-PE strategy achieves the best overall performance balance. Although the No-PE strategy obtains a marginally higher HSS (0.645) and an FB closer to the ideal value (0.964), this improvement comes at the cost of a elevated false alarm ratio (0.153). In contrast, the Mag-PE strategy achieves the highest TSS of 0.661 while suppressing the FAR to 0.132. This indicates that explicit spatial positional information enables the model to more precisely focus on critical structural features associated with flare triggering within magnetograms, thereby effectively filtering background noise while maintaining stable overall accuracy.

Conversely, when PE is applied to both branches (All-PE), model performance degrades compared to Mag-PE, with TSS dropping to 0.659 and FAR rising to 0.139. This phenomenon suggests that physical parameters, serving as highly condensed global statistics, inherently lack the spatial arrangement dependencies characteristic of image pixels. Imposing positional constraints on this modality may disrupt its original semantic structure and introduce redundant interference, thereby weakening the cross-modal feature fusion.

In summary, to achieve a balance between flare capture capability (high TSS) and warning reliability (low FAR), the Mag-PE strategy is adopted as the default positional encoding configuration in this work.

\subsubsection{Image Resolution}
\begin{table}
	\renewcommand{\arraystretch}{1.1}
	\setlength{\tabcolsep}{0pt} 
	\begin{threeparttable}  	
		\caption{\makebox[\linewidth][l]{Performance comparison of models using different input image resolutions.}}
		\label{tab11:Ab_resolution}
		\begin{tabular*}{\linewidth}{@{\extracolsep{\fill}}lccccccccc}
			\toprule
			Resolution & Mag. & Param. &  PE$_{\text{Mag}}$ & PE$_{\text{Param}}$ & HSS & TSS & FB & FAR & ACC \\
			\midrule
			$160 \times 160$ & $\cmark$ & $\cmark$ & $ \cmark $ & $\xmark$ & 0.613 & 0.653 & 0.876 & 0.123 & 0.825 \\ 
			$\mathbf{240 \times 240}$ 
			& $\cmark$ & $\cmark$ & $ \cmark $ & $\xmark$ & 0.658 & 0.661 & 0.909 & 0.132 & 0.829 \\
			$320 \times 320$ & $\cmark$  & $\cmark$ & $ \cmark $ & $\xmark$ & 0.621 & 0.660 & 0.884 & 0.122& 0.828 \\
			\bottomrule
		\end{tabular*}
		\begin{tablenotes}
			\item Same as in Table \ref{tab08:Ab_fusion}.
		\end{tablenotes}
	\end{threeparttable}    
\end{table}
To evaluate the impact of magnetogram resolution on model performance, we conduct a comparative analysis of three resolution settings $160 \times 160$, $240 \times 240$, and $320 \times 320$ pixels. The experimental dataset spans the observation period from May 1, 2010, to May 1, 2023, comprising $N$ ARs for training and $M$ ARs for testing. The results are summarized in Table \ref{tab11:Ab_resolution}.

The results indicate that the resolution to $160 \times 160$ leads to the loss of fine-grained textural details, thereby diminishing the predictive accuracy of the model. Conversely, increasing the magnetograms resolution to $320 \times 320$ yields no additional gains, suggesting that performance reaches saturation. This phenomenon is likely attributable to two primary factors: (1) computational constraints force a reduction in batch size, leading to prolonged training duration and potential optimization instability; and (2) the receptive field or parameter capacity of the current architecture appears insufficient to fully exploit fine-grained features beyond the $240 \times 240$ scale. Consequently, based on the trade-off between computational efficiency and predictive precision, we adopt the input resolution of magnetograms to $240 \times 240$ pixels.

\subsection{Feature Visualization}
\begin{figure*}  
	\centering
	\subfloat{
		\includegraphics[width=0.78\linewidth]{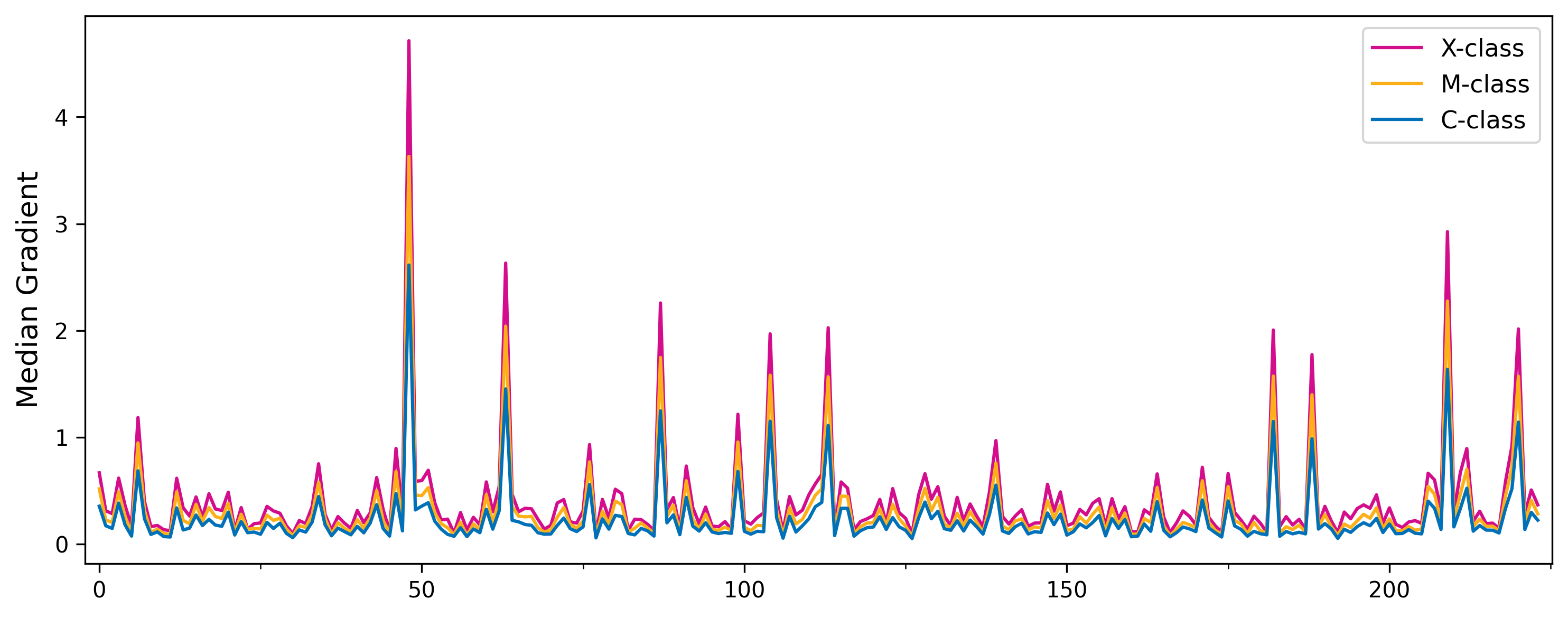}}\\
	\subfloat{
		\includegraphics[width=0.78\linewidth]{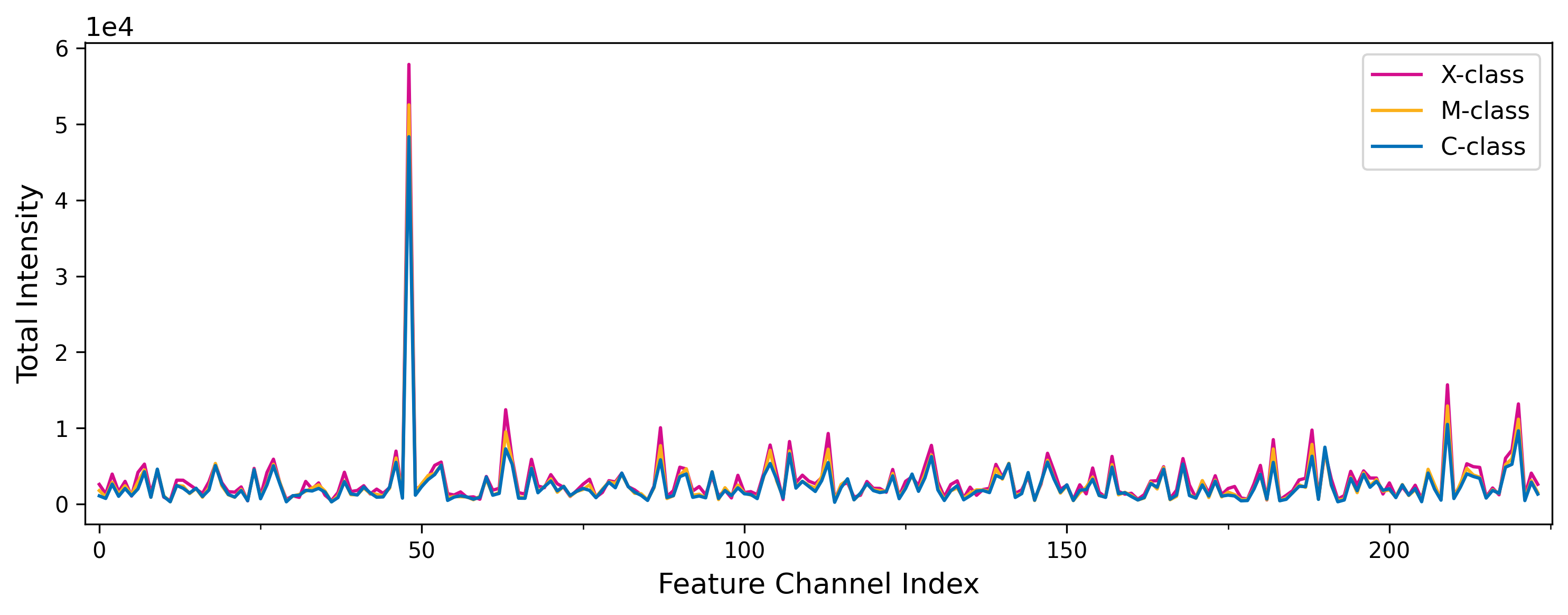}}\\
	\qquad 
	\caption{ 
		Statistical comparison of feature channel responses across flare classes. The upper and lower panels display the median gradient and total intensity of the feature extraction layer across channels, respectively. Curves of different colors correspond to different flare classes.}
	\label{fig09:gradient}	
\end{figure*}
To address the challenge of interpretability in deep learning models, we conduct a multi-dimensional investigation into the feature layers \citep[e.g.,][]{nishizuka2018deep,zheng2021hybrid}. In the domain of computer vision, the gradients of feature maps typically characterize texture variation rates or edge sharpness; in photospheric magnetograms, however, this corresponds directly to the intensity of magnetic field gradients,a critical physical quantity for flare eruptions \citep{falconer2011tool}. As shown in Figure~\ref{fig09:gradient}, we visualize the median gradient and total intensity across feature channels for different flare classes. The analysis reveals that the model does not treat all channels equally; instead, sharp response peaks emerged only in specific dominant channels.

Notably, in these key channels, the feature response intensity exhibits a strict hierarchical distribution (i.e., X-class $ > $ M-class $ > $ C-class).
This significant differentiation suggests that the model has autonomously learned a feature representation strategy, relying on a set of highly discriminative feature representations to identify complex magnetic structures within active regions, rather than weighting all feature channels uniformly.
Crucially, this strong activation in specific channels reflects the model's acute capture of sharp variations in magnetic structures. Consequently, it directly demonstrates that the learned feature extraction mechanism is highly consistent with the physical nature of energy release.

\begin{figure} 
	\centering
	\includegraphics[width=1.0\linewidth]{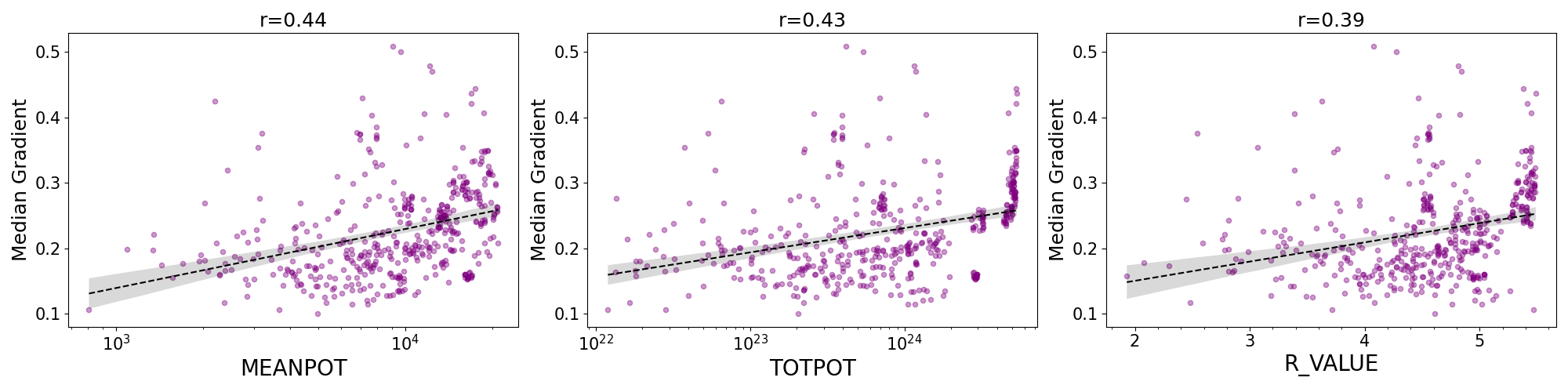}
	\caption{Scatter plots illustrate the relationships between the global network median gradient and MEANPOT, TOTPOT, and R\_VALUE. Black dashed lines indicate linear regression trends with confidence intervals denoted by gray shading. Pearson correlation coefficients are annotated at the top of each panel.}
	\label{fig10:scatter}
\end{figure}
To further quantify the connection between the internal representations of the model and physics mechanisms, we analyze the correlation between the network median gradient extracted by the model and the classic physical parameter. These parameters are MEANPOT, TOTPOT, and R\_VALUE \citep[e.g.,][]{bobra2014helioseismic,wang2019parameters}. As illustrated in Figure~\ref{fig10:scatter}, unlike channel-specific metrics, this value is derived by calculating the median of gradient magnitudes across the entire feature tensor which encompasses all spatial dimensions and channels. This global aggregation strategy ensures a holistic assessment of the topological complexity of the sample.

The results demonstrate a significant positive correlation between the network gradient response and all three physical parameters, with Pearson correlation coefficients of 0.44, 0.43, and 0.39 respectively. This finding indicates that the feature representations of the model are more strongly activated when the input magnetogram exhibits higher magnetic free energy density or stronger magnetic flux near the polarity inversion line \citep[e.g.,][]{leka2003photospheric,cicogna2021flare}. Such statistical consistency confirms that the internal feature extraction mechanism aligns closely with the physical nature of magnetic parameters \citep{ran2022relationship}.

\begin{figure} 
	\centering
	\includegraphics[width=0.9\linewidth]{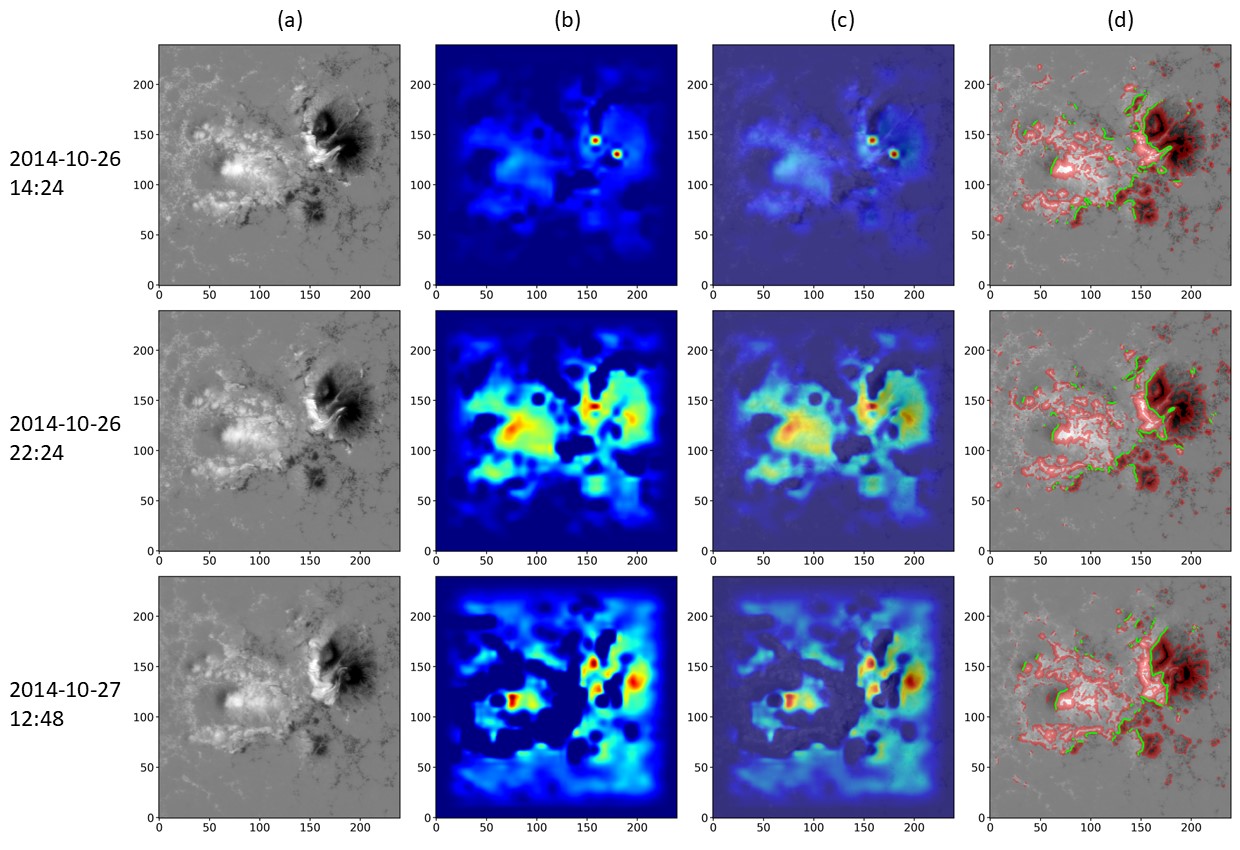}
	\caption{ Spatiotemporal evolution of model attention. Rows illustrate the evolution of model attention within the prediction window for an X-class flare event. Columns from left to right display: (a) the original line-of-sight magnetogram; (b) the model attention heatmap where warm colors indicate high activation; (c) the overlay of the magnetogram and heatmap; and (d) the magnetogram superimposed with features where red shading marks strong magnetic gradient regions and green contours delineate the strong-gradient polarity inversion line. Each block is 240 $\times$ 240 pixels in size.}
	\label{fig11:hotmap_X}	
\end{figure}
\begin{figure} 
	\centering
	\includegraphics[width=0.9\linewidth]{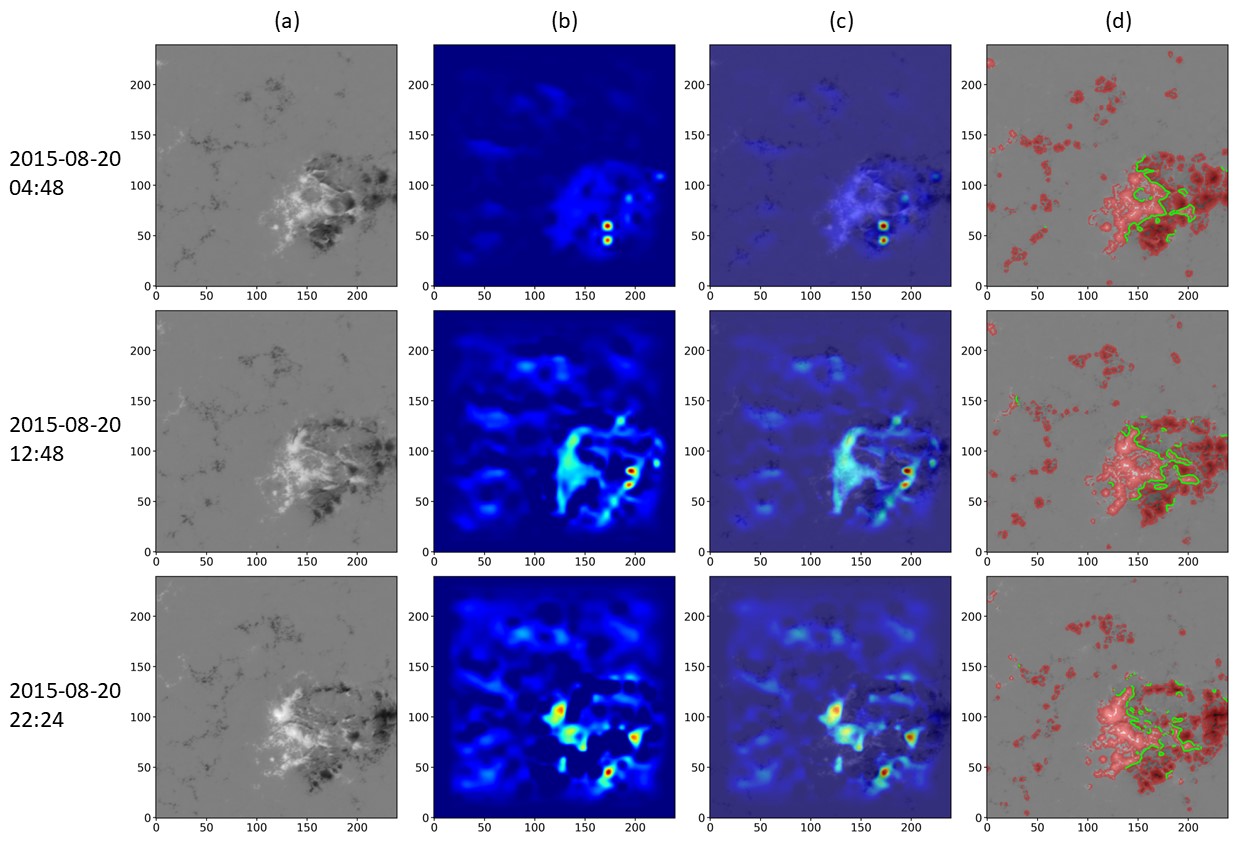}
	\caption{Same as Figure~\ref{fig11:hotmap_X}, but for an M-class flare event.}
	\label{fig12:hotmap_M}
\end{figure}
\begin{figure} 
	\centering
	\includegraphics[width=0.9\linewidth]{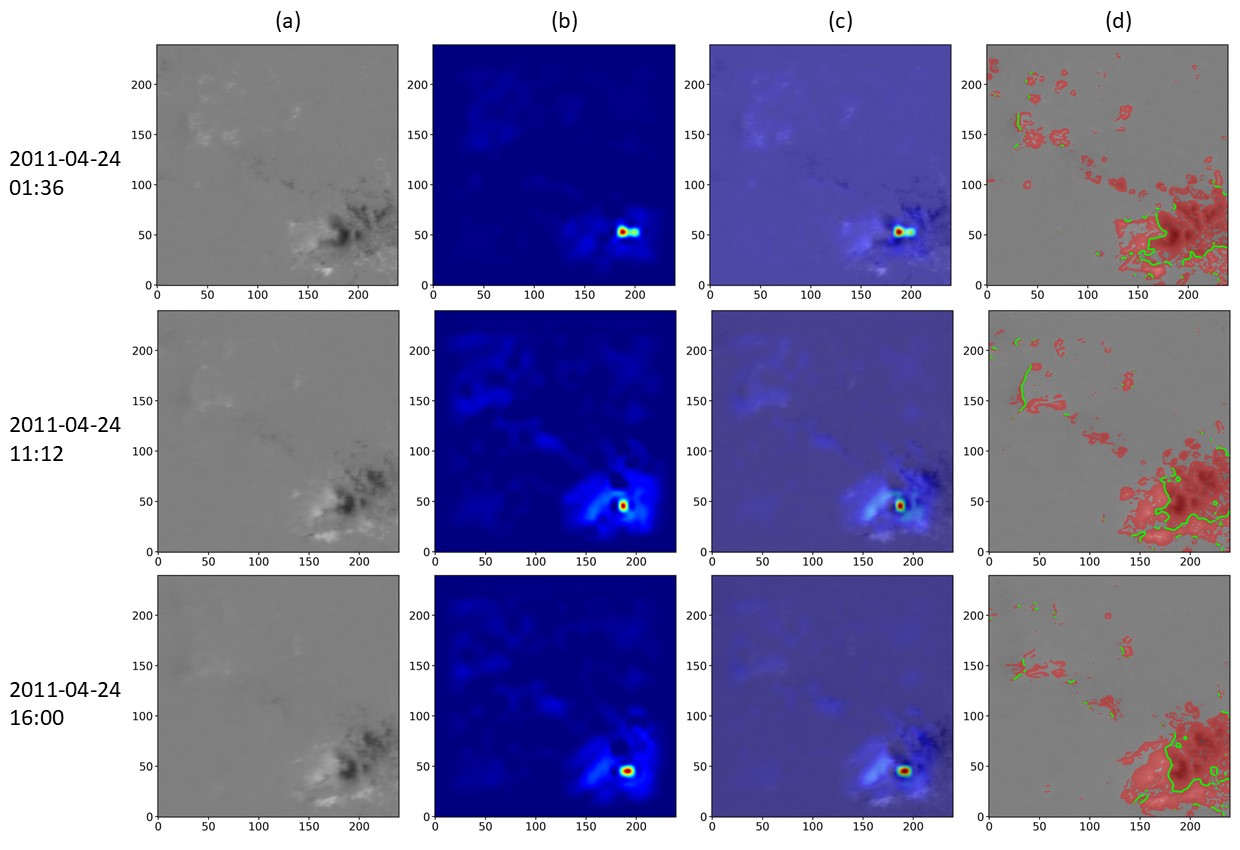}
	\caption{Same as Figure~\ref{fig11:hotmap_X}, but for a C-class flare event.}
	\label{fig13:hotmap_C}
\end{figure}
To verify whether the model spatially locks onto the physical trigger source of eruptions, we visualiz the attention maps generated by the model. We then compare them spatially with the calculated strong vertical magnetic field gradient and the strong-gradient polarity inversion line \citep[e.g.,][]{yi2021visual,cicogna2021flare}. Figures~\ref{fig11:hotmap_X}-\ref{fig13:hotmap_C} illustrate the spatiotemporal trajectory of the attention evolution for different flare classes.

As shown in the first row of Figures \ref{fig11:hotmap_X}-\ref{fig13:hotmap_C}, the attention map exhibits low-intensity activation features characterized by cool colors, with a relatively divergent focus. Consistent with this physical state, the magnetic topology of the corresponding active region is relatively loose, and the polarity inversion line presents a fragmented morphology. This indicates that the model maintains a low alert level at this stage.

The second row reveals a distinct shift in the perception of the model as the attention focus automatically converges on the central region where positive and negative magnetic poles approach each other. At this moment, the area of high-gradient regions indicated by red shading expands, and the polarity inversion line begins to elongate and complicate, signaling the injection of non-potential energy \citep{yang2013magnetic}. This synchronous enhancement proves that the model keenly captures the dynamic process of flux emergence and responds in real-time to the accumulation of magnetic stress \citep{korsos2015flare}.

In the third row, the active region displays a highly non-potential configuration where a compact high-gradient region tightly wraps around a twisted polarity inversion line. The high-intensity hotspot in the attention map spatially coincides precisely with the physical strong-gradient polarity inversion line. This high degree of morphological consistency confirms that the model has successfully locked onto the physical core region responsible for the eruption.

In summary, the evolutionary process from dispersed low activation to focused high activation provides compelling evidence that the proposed model is not a black box. Instead, it effectively integrates local magnetogram textures into semantic physical precursors, thereby establishing its physical credibility in solar flare forecasting.

\section{Discussions and Conclusions}
\label{sec5:Conclusions}
This study demonstrates the efficacy of a dual-branch multimodal fusion model for 24-hour solar flare prediction. The analysis encompasses binary classification and extends to a multi-class task targeting fine-grained C, M, and X-class flare intensities. To strictly evaluate the model's robustness, we implement a stratified group five-fold cross-validation strategy across five constructed subsets (F1–F5), reporting the average metrics to demonstrate robust generalization.

The reliability of solar flare prediction models depends heavily on the integrity of the data splitting and evaluation strategy. In this study, we emphasize the necessity of the splitting-before-sampling strategy, which fundamentally departs from the sampling-before-splitting approach common in many previous studies \citep[e.g.,][]{zheng2019solar, zheng2021hybrid}. By ensuring that all class-balancing operations are performed independently within each partition only after the split is finalized, we eliminate potential statistical coupling between training and testing data. This rigor is essential for providing an unbiased assessment of a model's true generalization capability on unseen solar active regions.

Furthermore, our dual-dataset evaluation (D1 and D2) for binary tasks offers a more comprehensive perspective than traditional balanced-only evaluations. While the balanced dataset D2 provides a controlled environment to assess the model's sensitivity to flare features, the original-distribution dataset D1 serves as a critical operational benchmark. By maintaining the inherent class imbalance of real-world observations in D1, we can better estimate the model's performance in practical forecasting scenarios, where the overwhelming majority of samples are non-flaring.

For multi-class discrimination, the adoption of stratified group five-fold cross-validation addresses the statistical instability inherent in rare event prediction. Unlike standard k-fold or year-based splitting \citep[e.g.,][]{li2020predicting, nishizuka2021operational}, our stratified approach ensures that extremely rare M and X-class flares are proportionally represented in every fold. This prevents the empty class problem in testing partitions and ensures that the reported average metrics across F1–F5 are statistically robust. Ultimately, this framework provides a rigorous standard for evaluating flare intensity discrimination, ensuring that the model learns feature representation rather than memorizing region-specific evolutionary signatures or benefiting from information leakage.

Among existing approaches, prior work such as \citet{nishizuka2018deep} relies on manual feature selection based on domain knowledge, which limits the ability to capture potentially critical features and reduces adaptability to changes in data distribution. Furthermore, some studies use single-modal inputs, utilizing either magnetograms \citep[e.g.,][]{huang2018deep,li2020predicting,zheng2023multiclass} or magnetic parameters \citep[e.g.,][]{zheng2023comparative,li2024prediction}, thereby overlooking the complementary information inherent across modalities. Although \citet{tang2021solar} incorporated multiple architectures and multimodal data, fusion is performed only at the decision level, resulting in limited interaction. In addition, while \citet{zheng2023multiclass} compared magnetogram-based and parameter-based models on a unified dataset, their physical model relied solely on time-series magnetic parameters, and the magnetogram-based model employed only single-scale magnetograms. These simplified designs neither enable effective complementary fusion between magnetogram and parameter modalities nor exploit multi-scale feature representations.  

In contrast, the proposed dual-branch multimodal fusion model separately extracts features from magnetograms and magnetic parameters integrated with cross-attention mechanisms. Subsequently, we introduce a cross-scale interactions mechanism to facilitate multi-scale feature fusion. Trained in an end-to-end manner, the model significantly enhances both feature representation capability and generalization performance. The consistent superiority of our model, particularly in predicting X-class flares, validates the efficacy of the multimodal fusion framework. As shown in the ablation study in Table \ref{tab08:Ab_fusion}, the integration of the physical parameter branch not only enhances prediction accuracy but, more critically, suppresses the False Alarm Rate, thereby boosting overall performance metrics.

This improvement underscores the intrinsic complementary mechanism between the two modalities. While networks are adept at extracting implicit spatial patterns from magnetograms, their sensitivity to pixel-level details renders them susceptible to interference from projection effects and high-frequency observational noise. Addressing this, SHARP parameters serve as a vital supplement. By providing a condensed representation of magnetic properties—such as total magnetic flux and current helicity—these parameters directly characterize the topological complexity and non-potential energy accumulation of active regions. Consequently, incorporating these strong physical priors acts as a form of regularization, making the model robust against local image perturbations. Fundamentally, this fusion strategy anchors the decision-making process to the core physics of flare eruptions rather than mere image pattern matching, effectively preventing the network from overfitting to superficial morphological features.

The main results of this work are summarized as follows. (1) We propose a dual-branch framework that utilizes multimodal and cross-scale interaction mechanisms to deeply fuse magnetograms with magnetic parameters. This end-to-end architecture effectively integrates complementary spatial and physical features, significantly enhancing representation learning and performance of solar flare prediction compared to single-modal designs. (2) This study introduces a stratified group five-fold cross-validation scheme and adopts a splitting-before-sampling strategy. The scheme ensures non-overlapping data sources between training and testing sets to prevent data leakage, maintains class proportions consistent with the overall dataset across all folds, and guarantees the presence of representative samples from all classes in both subsets. (3) The model demonstrates superior predictive performance on X-class flares, significantly outperforming C and M-class predictions. Despite the scarcity of X-class samples, the model achieves high TSS and HSS scores while maintaining a low false alarm rate, surpassing several existing methods. These results indicate that the proposed multimodal fusion architecture possesses a distinct advantage in capturing the discriminative features of strong solar flare eruptions.
 
In conclusion, while the proposed model shows robust performance on X-class and C-class flares prediction, reducing the false positive rate for M-class flares remains an important task. In future work, we aim to advance beyond the current framework by designing a spatiotemporal prediction model that integrates time-series physical evolution with magnetogram evolution. Specifically, we plan to incorporate physics-informed semantic guidance to better direct the magnetogram learning process. Additionally, we will focus on optimizing the feature extraction mechanism and introducing strategies such as class-aware loss functions or dynamic adjustments. These enhancements are expected to resolve the ambiguity in M-class events and further elevate the model's overall discriminative performance.

\begin{acknowledgments}
We sincerely thank the reviewers for their thorough evaluation of our manuscript and for their valuable comments and suggestions. The data used in this study are provided by NASA/SDO, the HMI science team, and the GOES team. We gratefully acknowledge the support of all members of the Joint Science Operations Center (JSOC) at Stanford University and thank them for granting us access to the relevant data. We also express our sincere appreciation to the SDO/HMI team members for their dedicated efforts and significant contributions to the SDO mission. This work is supported by the National Key R\&D Program of China (2021YFA1600500, 2025YFF0510700, 2025YFE0202500, 2022YFF0503800), the National Natural Science Foundation of China (42274201) and the Specialized Research Fund for State Key Laboratory of Solar Activity and Space Weather.

\end{acknowledgments}

\begin{contribution}
Limin Zhao proposed the methodology, performed the experiments, and was responsible for writing and submitting the manuscript. Yihua Yan provided overall supervision and critical input on complex aspects of the research. All other authors contributed to the revision through constructive feedback and helped improve the manuscript.

%
%
\end{contribution}

\bibliography{sample701}{}
\bibliographystyle{aasjournalv7}



\end{document}